\documentclass[onecolumn,a4paper]{IEEEtran}
\usepackage{amsmath,amsfonts,amssymb,bm,mathrsfs}
\usepackage{color}
\usepackage{comment}
\usepackage{cite}
\usepackage{graphicx}
\newcommand{\sh}{\mathsf{h}}
\newcommand{\sk}{\mathsf{k}}
\newcommand{\sx}{\mathsf{x}}
\newcommand{\sy}{\mathsf{y}}
\newcommand{\sbx}{\bm{\mathsf{x}}}
\newcommand{\sby}{\bm{\mathsf{y}}}

\begin{document}

\author{Fran\c{c}ois Vernotte, Michel Lenczner, Pierre-Yves Bourgeois, and Enrico Rubiola%
\thanks{FV is with UTINAM, Observatory THETA, University of Franche-Comt\'{e} / UBFC / CNRS, 41 bis avenue de l'Observatoire - B.P. 1615, 25010 Besan\c{c}on Cedex, France (Email: \texttt{francois.vernotte@obs-besancon.fr}).} \thanks{ML, PYB and ER are with the CNRS FEMTO-ST Institute, Dept.\ of Time and Frequency (UBFC, UFC, UTBM, ENSMM), 26, chemin de l'\'{E}pitaphe, Besancon, France. E-mail \{michel.lenczner$|$pyb2$|$rubiola\}@femto-st.fr. 
Enrico's home page http://rubiola.org.}%
\\[1em] 
\bfseries\sffamily
IEEE Transact.\ UFFC Special Issue to celebrate the\\[0.5ex]
{\Large 50th anniversary of the Allan Variance}}
\title{The Parabolic Variance (PVAR), a Wavelet Variance Based on the Least-Square Fit}
\date{\today}
\maketitle


\begin{abstract}

This article introduces the Parabolic Variance (PVAR), a wavelet variance similar to the Allan variance, based on the Linear Regression (LR) of phase data.  The companion article\footnote{The \emph{companion article} arXiv:1506.05009 [physics.ins-det] has been submitted for the 2015 IEEE IFCS Special Issue of the IEEE Transact.\ UFFC.}  arXiv:1506.05009 [physics.ins-det] details the $\Omega$ frequency counter, which implements the LR estimate.

The PVAR combines the advantages of AVAR and MVAR\@.  PVAR is good for long-term analysis because the wavelet spans over $2\tau$, the same of the AVAR wavelet; and good for short-term analysis because the response to  white and flicker PM is $1/\tau^3$ and $1/\tau^2$, same as the MVAR\@.

After setting the theoretical framework, we study the degrees of freedom and the confidence interval for the most common noise types.  Then, we focus on the detection of a weak noise process at the transition -- or \emph{corner} -- where a faster process rolls off.  This new perspective raises the question of which variance detects the weak process with the shortest data record. 
Our simulations show that PVAR is a fortunate tradeoff.
PVAR is superior to MVAR in all cases, exhibits the best ability to divide between fast noise phenomena (up to flicker FM), and is almost as good as AVAR for the detection of random walk and drift.

\end{abstract}

\section{Introduction}
The Allan variance (AVAR) was the first of the wavelet-like variances used for the characterization of oscillators and frequency standards \cite{allan1966}.  After 50 years of research, AVAR is still unsurpassed at rendering the largest $\tau$ for a given time series of experimental data.  This feature is highly desired for monitoring the frequency standards used for timekeeping.   

Unfortunately, AVAR is not a good choice in the region of fast noise processes.  In fact, the AVAR response to white and flicker PM noise is nearly the same, $1/\tau^2$.  For short-term analysis, other wavelet variances are preferred, chiefly the modified Allan variance (MVAR) \cite{Snyder-1980-AO--Mod-Allan, Allan-Barnes-1981-FCS--Mod-Allan, Snyder-1981-IFCS}.  The MVAR response is $1/\tau^3$ and $1/\tau^2$ for white and flicker PM, respectively.  However, MVAR is poor for slow phenomena because the wavelet spans over $3\tau$ instead of $2\tau$.  Thus, for a data record of duration $T$, the absolute maximum $\tau$ is $T/3$ instead of $T/2$.

Speaking of `wavelet-like' variances, we review the fundamentals.  A wavelet $\psi(t)$ is a shock with energy equal to one and average equal to zero, whose energy is well confined in a time interval (see for example \cite[p.~2]{percival:wavelets}) called `support' in proper mathematical terms.  In formula, $\int_\mathbb{R}\psi^2(t)\:dt=1$, $\int_\mathbb{R}\psi(t)\:dt=0$, and $\int_{-a/2}^{a/2}\psi^2(t)\:dt=1-\epsilon$, with small $\epsilon>0$.
It makes sense to re-normalize the wavelet as $\frac{1}{a}\int_\mathbb{R}\psi^2(t)\:dt=1$, so that it is suitable to power-type signals (finite power) instead of energy-type signals (energy finite). By obvious analogy, we use the terms `power-type wavelet' and `energy-type wavelet'.
These two normalizations often go together in spectral analysis and telecom (see the classical books \cite{Papoulis-1962--Fourier,Proakis:comm2ed}).  
For historical reasons, in clock analysis we add a trivial coefficient that sets the response to a linear drift $\mathsf{D_y}$ to $\frac{1}{2}\mathsf{D_y}^2$, the same for all the variances.

High resolution in the presence of white and flicker phase noise is mandatory for the measurement of short-term fluctuations ($\mu$s to s), and useful for medium-term fluctuations (up to days). This is the case of optics and of the generation of pure microwaves from optics. The same features are of paramount importance for radars, VLBI, geodesy, space communications, etc. As a fringe benefit, extending the time-domain measurements to lower $\tau$ is useful to check on the consistency between variances and phase noise spectra.  MVAR is suitable to the analysis of fast fluctuations, at a moderate cost in terms of computing power.  Frequency counters specialized for MVAR are available as a niche product, chiefly intended for research labs \cite{kramer2001}.  

A sampling rate of $1/\tau$ is sufficient for the measurement of AVAR, while a rate of $1/\tau_0=m/\tau$ is needed for MVAR, where the rejection of white phase noise is proportional to $m$.  MVAR is based on the simple averaging of $m$ fully-overlapped (spaced by the sampling step $\tau_0$) frequency data, before evaluating $\sigma^2(\tau)$.

The linear regression provides the lowest-energy (or lowest-power) fit of a data set, which is considered in most cases as the optimal approximation, at least for white noise.  For our purposes, the least-square fit finds an obvious application in the estimation of frequency from a time series of phase data, and opens the way to improvements in fluctuation analysis.  Besides, new digital hardware --- like Field-Programmable Gate Arrays (FPGAs) and Systems on Chip (SoCs) --- provides bandwidth and computing power at an acceptable complexity, and makes possible least-square fitting in real-time.

We apply least-square estimation of frequency to fast time stamping. The simplest estimator in this family is the linear regression (LR) on phase data. The LR can be interpreted as a weight function applied to the measured frequency fluctuations. The shape of such weight function is parabolic.  The corresponding instrument is called `$\Omega$ counter,' described in the companion article \cite{rubiola2015omega}.  The name $\Omega$ comes from the graphical analogy of the parabola with the Greek letter, in the continuity of the $\Pi$ and $\Lambda$ counters \cite{rubiola2005rsi, dawkins2007}.  The $\Omega$ estimator is similar to the $\Lambda$ estimator, but exhibits higher rejection of the instrument noise, chiefly of white phase noise. This is important in the measurement of fast phenomena, where the cutoff frequency $f_H$ is necessarily high, and the white phase noise is integrated over the wide analog bandwidth that follows.

In the same way as the $\Pi$ and $\Lambda$ estimators yield the AVAR and the MVAR, respectively, we define a variance based on the $\Omega$ estimator.  Like in the AVAR and MVAR, the weight functions are similar to wavelets, but for the trivial difference that they are normalized for power-type signals.  A similar use of the LR was proposed independently by Benkler et al.\ \cite{benkler2015} at the IFCS, where we gave our first presentation on the $\Omega$ counter and on our standpoint about the PVAR\@.  In a private discussion, we agreed on the name PVAR (Parabolic VARiance) for this variance, superseding earlier terms \cite{benkler2015discussion}.

We stress that the wavelet variances are mathematical tools to describe the frequency stability of an oscillator (or the fluctuation of any physical quantity).  Albeit they have similar properties, none of them should be taken as ``the stability'' of an oscillator.  For the same reason, MVAR and PVAR should not be mistaken as `estimators' of the AVAR\@.  To this extent, the only privilege of AVAR is the emphasis it is given in standard documents \cite{IEEE-STD-1139-2008}.

After setting the theoretical framework of the PVAR, we provide the response to noise described by the usual polynomial spectrum.  Then we calculate the degrees of freedom and confidence intervals, checking on the analytical results against extensive simulations.  Finally, we compare the performance of AVAR, MVAR and PVAR for the detection of noise types, using the value of $\tau$ where $\sigma^2(\tau)$ changes law as an indicator.  In most practical cases PVAR turns out to be the fastest, to the extent that it enables such detection with the shortest data record.

\section{Statement of the Problem}

\noindent The clock signal is usually written as
\begin{align*}
v(t)=V_0 \sin[2\pi\nu_0t+\varphi(t)]
\end{align*}
where $V_0$ is the amplitude, $\nu_0$ is the nominal frequency, and $%
\varphi(t)$ is the random phase fluctuation. Notice that $\varphi(t)$ is
allowed to exceed $\pm\pi$. Alternatively, randomness is ascribed to the
frequency fluctuation $(\Delta\nu)(t)=2\pi\dot{\varphi}(t)$.

We introduce the normalized quantities
\begin{align*}
\sbx(t)&=t+\sx(t) \\
\sby(t)&=1+\sy(t)
\end{align*}
where $\sx(t)=\varphi(t)/2\pi\nu_0$, and $\sby(t)=\dot{\sbx}(t)$. The quantity $\sbx(t)$ is the clock
readout, which is equal to the time $t$ plus the random fluctuation $\mathsf{%
x}(t)$. Accordingly, the clock signal reads
\begin{align*}
v(t)&=V_0 \sin[2\pi\nu_0\sbx(t)] \\
&=V_0 \sin[2\pi\nu_0t+2\pi\nu_0\sx(t)]
\end{align*}
For the layman, $\sbx$ is the time displayed by a watch, $t$ is
the `exact' time from a radio broadcast, and $\sx$ the watch error.
The error $\sx$ is positive (negative) when the watch leads (lags).
Similarly, $\sby$ is the normalized frequency of the watch's
internal quartz, and $\sy$ its fractional error. For example, if $\sy =
+10$ ppm (constant), the watch leads uniformly by 1.15 s/day. For the
scientist, $\sx(t)$ is the random time fluctuation, often referred to
as `phase time' (fluctuation), and $\sy(t)$ is the random
fractional-frequency fluctuation. The quantities $\sx(t)$ and $%
\sy(t)$ match exactly $x(t)$ and $y(t)$ used in the general
literature of time and frequency \cite{Barnes-et-al-1971-TIM--Frequency-stability,Vanier-Audoin-1989--Frequency-standards,IEEE-STD-1139-2008}

\begin{figure}[t]\centering
\includegraphics[width=8cm]{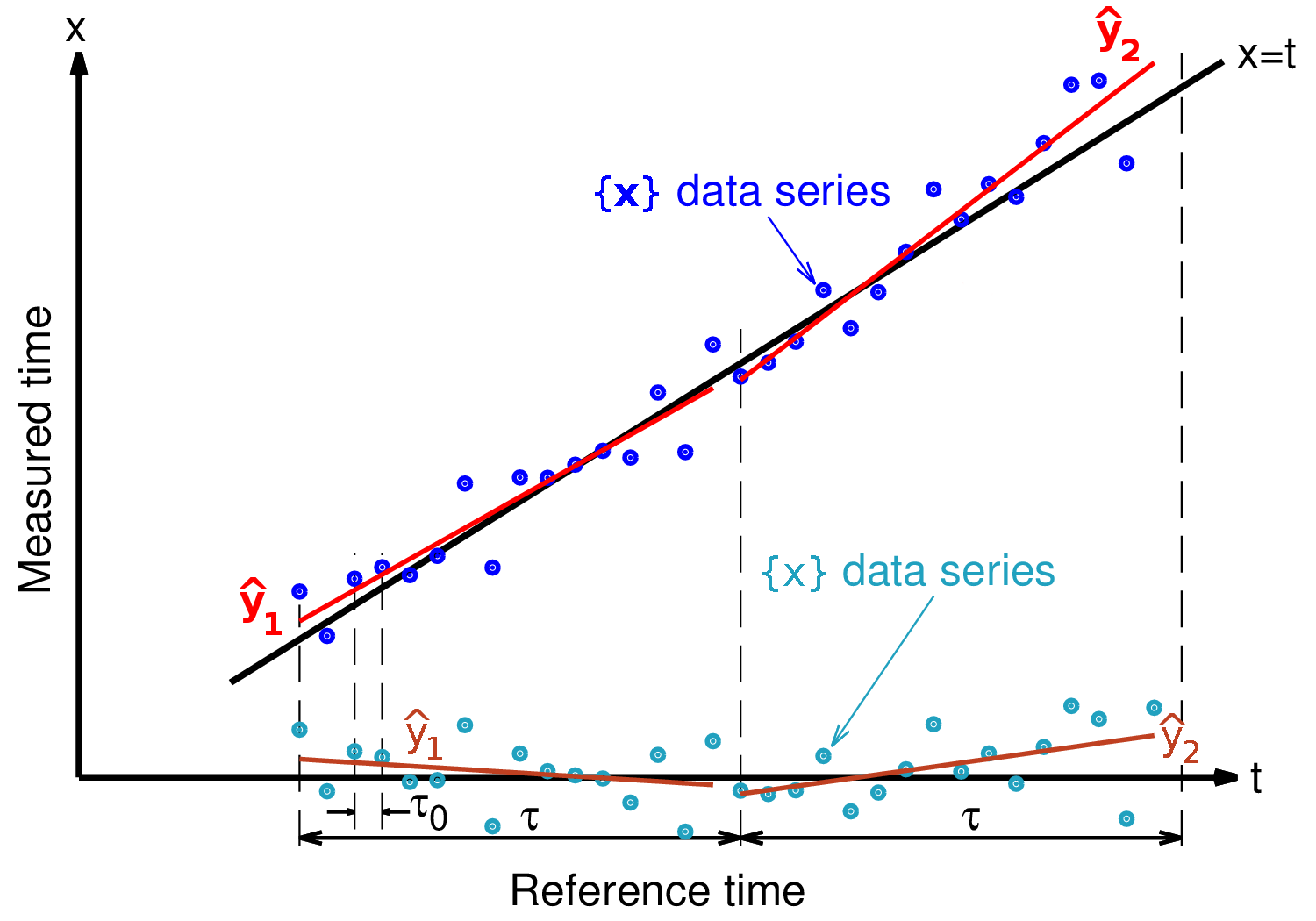}
\caption{Principle of two-sample linear regression measurement, and notation.} 
\label{fig:notation}
\end{figure}

The 
 main point
 of this article is explained in Fig.~\ref{fig:notation}. We use the
linear regression of phase data to get a sequence $\{\hat{\sby}\}$
of data averaged on contiguous time intervals of duration $\tau$, and in
turn the sequence $\{\hat{\sy}\}$ of fractional-frequency fluctuation
data. Two contiguous elements of $\{\hat{\sby}\}$ and $\{\hat{%
\sy}\}$ are shown in Fig.~\ref{fig:notation}, from which we get one
value of $\frac{1}{2}(\sy_2-\sy_1)^2$ for the estimation of
 the variance.

Most of the concepts 
 below 
 are expressed in both the continuous and the discrete settings with common notations without risk of confusion. For
example, the same expression $\sbx =t+\sx$ maps into 
$\sbx_{k}=t_{k}+\sx_{k}$ in the discrete case, and into $%
\sbx(t)=t+\sx(t)$ in the continuous case. The notations $%
\left\langle\,.\,\right\rangle$, $(.,.)$ and $||\,.\,||$ represent the
average, the scalar product and the norm. They are defined as $\left\langle %
\sbx\right\rangle =\frac{1}{n}\sum_{k}\sbx_{k},$ $(%
\sbx,\sby )=\frac{1}{n}\sum_{k}\sbx_{k}%
\sby_{k}$, $||\sbx||=(\sbx,\sbx%
)^{1/2}$ where $n$ is the number of terms of the sum in the discrete case,
and as $\left\langle \sbx\right\rangle =\frac{1}{T}\int %
\sbx(t)$ $dt,$ $(\sbx,\sby)=\frac{1}{T}\int %
\sbx(t)\sby(t)$ $dt$, $||\sbx||=(%
\sbx,\sbx)^{1/2}$ where $T$ is the length of the
interval of integration in the continuous case. The span of the sum and the
integral will be made precise in each case of application. The mathematical
expectation and the variance of random variables are denoted by $\mathbb{E}%
\{\,.\,\}$ and $\mathbb{V}\{\,.\,\}$.

The linear regression problem consists in searching the optimum value $\hat{\sby}$ of the slope $\eta$ (dummy variable) that minimizes the norm of the error $\sbx-\sbx_0-\eta t$, i.e., $\hat{\sby}=\arg\min_{\eta}||\sbx-\sbx_0-\eta t||^{2}$.
Since we are not interessed in $\sbx_0$, which only reflects choice of the origin of $\sbx$, the solution is the random variable
\begin{equation*}
\hat{\sby}=\frac{\left( \sbx-\left\langle %
\sbx\right\rangle ,t-\left\langle t\right\rangle \right) }{%
||t-\left\langle t\right\rangle ||^{2}}.
\end{equation*}
We recall some useful properties of $\hat{\sby}$ as an estimator
of the slope of $\sbx$. For the sake of simplicity, with no loss of
generality, we refer to a time sequence is centered at zero, i.e., $\left<t\right>=0$.

\begin{enumerate}
\item The estimator $\hat{\sby}$ can be simplified as
\begin{equation*}
\hat{\sby}=\frac{\left( \sbx,t\right) }{||t||^{2}}.
\end{equation*}

\item If the component $\sx_k$ (or the values $\sx(t)$) are independent, the estimator variance is 
\begin{equation*}
\mathbb{V}\{\hat{\sby}\}=\frac{\sigma_{\sx}^{2}}{||t||^{2}}
\end{equation*}

The assumption of independent continuous random process is rather usual in theoretical works. However this is done to simplify some proofs, the results can be used in their discrete form.

\item Sampling uniformly at the interval $\tau_0$, the discrete time is $t_{k}=(k+\frac{1}{2})\tau _{0}$ for $k\in \{-p,...,p\}$, $m=2p$ and $\tau=m\tau _{0}$. For large $m$, we get
\begin{equation*}
\hat{\sby}\approx 1+\frac{12\left( \sx,t\right) }{m\tau^{2}}
\qquad\mathrm{and}\qquad
\mathbb{V}\{\hat{\sby}\}\approx\frac{12\sigma_{\sx}^{2}}{m\tau ^{2}}.
\end{equation*}

\item With a signal that is continuous over a symmetric time interval $(-%
\frac{\tau }{2},\frac{\tau }{2})$, we get
\begin{equation}
\hat{\sby}=1+\frac{12\left( \sx,t\right) }{\tau ^{3}}.
\label{LR estimator - continuous case}
\end{equation}
\end{enumerate}
The continuous form of the estimator $\hat{\sby}$ can be expressed as a weighted average of $\sbx$ or $\sby$. For this purpose, it is useful to take $\hat{\sby}$ as a time dependent function defined over $t\in (0,\tau)$
\begin{eqnarray}
\hat{\sby}(t) 
&=&\frac{12}{\tau^{3}}\int_{-\tau/2}^{\tau /2}s\:\sbx(t-\tau/2+s)\;ds \nonumber\\
&=&\frac{12}{\tau^{3}}\int_{-\tau}^{0}(s+\tau/2)\:\sbx(t+s)\;ds \nonumber\\
&=&\frac{12}{\tau^{3}}\int_{0}^{\tau}(\tau/2-s)\:\sbx(t-s)\;ds.  \label{eq:yest}
\end{eqnarray}


\section{Time Domain Representation}
\subsection{Generic Wavelet Variance}
Let us denote with $T$ the duration of the data run, with $\tau _{0}$ the sampling interval, with $N$ the number of samples, and with $n$ the ratio $T/\tau$.  Thus, $T=N\tau _{0}$, and $N=mn$.  We consider the series $\{\hat{\sy}_{i}\}_{i=1,..,n}$ of frequency deviation estimates.
In this section we denote with $\sigma^2(\tau)$ a generic wavelet variance, either AVAR, MVAR, PVAR, etc.

In the case of uncorrelated frequency fluctuations (white FM),
an unbiased estimator of the variance 
$\mathbb{V}\left\{ \hat{\sy}\right\} $ is 
\begin{equation*}
S_{n-1}^{2}=\frac{1}{n-1}\sum_{i=1}^{n}\Bigl(\hat{\sy}_{i}-\frac{1}{n}\sum_{j=1}^{n}\hat{\sy}_{j}\Bigr)^{2},
\end{equation*}
so
\begin{equation*}
\mathbb{V}\left\{ \hat{\sy}\right\} =\mathbb{E}\left\{S_{n-1}^{2}\right\} .
\end{equation*}
After Allan \cite{allan1966}, we replace the estimator $S_{n-1}^{2}$ with a two-sample variance by setting $n=2$.
Then, the variance $\mathbb{V}\left\{\hat{\sy}\right\}=\mathbb{E}\left\{S_1^2\right\}$ is 
\begin{equation}
\sigma ^{2}(\tau )=\frac{1}{2}\mathbb{E}\left\{\left(\hat{\sy}_1-
\hat{\sy}_2\right) ^{2}\right\},  \label{eq:allan}
\end{equation}%
and its estimator averaged over the $n-1$ terms 
\begin{equation}
\hat{\sigma }^{2}(\tau )=\frac{1}{2}\left\langle \left( \hat{\sy}%
_{i}-\hat{\sy}_{i+1}\right) ^{2}\right\rangle.
\label{eq:estimator-of-allan}
\end{equation}
Notice that two-sample variance is generally written as $\sigma_\sy^2(\tau)$, and that we drop the subscript $\sy$.

Following the Lesage-Audoin approach \cite{lesage1973}, we define the point variance estimates
\begin{equation}
\alpha _{i}=\frac{1}{\sqrt{2}}\bigl(\hat{\sy}_{i}-\hat{\sy}_{i+1}\bigr)
\end{equation}
and the estimated variance
\begin{equation}
\hat{\sigma}^2(\tau )=\frac{1}{M}\sum_{i=1}^{M}\alpha_i^2.
\label{eq:var_gen}
\end{equation}
The relationship between the $\alpha_{i}$ and the $N$ individual $\sx(k\tau _{0})$ measures depends on the type of counter ($\Pi$, $\Lambda$, $\Omega$).

\subsection{Continuous-Time Formulation of PVAR} 
In the case of continuous time, the difference between contiguous measures is 
\begin{eqnarray*}
\hat{\sy}(t+\tau )-\hat{\sy}(t) 
&=&\frac{12}{\tau ^{3}}
\left[\int_{0}^{\tau}\Bigl(\frac{\tau}{2}-s\Bigr)\:\sx(t+\tau -s)\;
ds-\int_{0}^{\tau}\Bigl(\frac{\tau}{2}-s\Bigr)\:\sx(t-s)\;ds\right] \\
&=&\frac{12}{\tau^{3}}\int_{-\tau}^{\tau}\Bigl(|s|-\frac{\tau }{2}\Bigr)\:\sx(t-s)\;ds \\
&=&\frac{12}{\tau^{3}}\int_{-\tau}^{\tau}\Bigl(|t-s|-\frac{\tau}{2}\Bigr)\:\sx(s)\;ds.
\end{eqnarray*}
Accordingly, the two-sample variance (\ref{eq:allan}) is written as
\begin{equation*}
\sigma_{P}^{2}(\tau)=\frac{1}{2}\mathbb{E}\Bigl\{\bigl(\hat{\sy}(t+\tau)-\hat{\sy}(t)\bigr)^{2}\Bigr\},
\end{equation*}
and notice the subscript $P$ for PVAR.  Such variance is independent of $t$, and it can be expressed as the running average
\begin{equation}
\sigma _{P}^{2}(\tau )=\mathbb{E}\Biggl\{ \Biggl(\int_{-\tau }^{+\tau }
\sx(s)\,w_{\sx}(s-t)\;ds\Biggr)^{2}\Biggr\},
\label{eq:PVAR}
\end{equation}
where 
\begin{equation*}
w_{\sx}(t)=\frac{6\sqrt{2}}{\tau ^{3}}\Bigl(|t|-\frac{\tau }{2}\Bigr)\:\chi_{(-\tau ,\tau )}(t)
\end{equation*}
is the even weight function, and
\begin{equation*}
\chi _{(-\tau,\tau)}(t)=\begin{cases}
1 & t\in(-\tau,\tau)\\
0 & \text{elsewhere}
\end{cases} 
\end{equation*}
is the indicator function (or characteristic function).  

From (\ref{eq:PVAR}), we see that PVAR can also be written as a convolution product
\begin{equation*}
\sigma_{P}^{2}(\tau)=\mathbb{E}\Biggl\{ \Biggl(\int_{-\infty }^{+\infty }
\sx(s)\,\mathfrak{h}_{\sx}(t-s)\;ds\Biggr)^{2}\Biggr\}=\mathbb{E}\Biggl\{\Biggl(\sx(t) \ast \mathfrak{h}_{\sx}(t)\Biggr)^{2}\Biggr\},
\end{equation*}
where $\mathfrak{h}(t)$ is the convolution kernel which applies to $\sx(t)$.
The kernel $\mathfrak{h}(t)$ is related to the weight function $w(t)$ by the general property that $\mathfrak{h}(t)=w(-t)$.  However, since $w_{\sx}(t)$, is even function, it holds that $\mathfrak{h}_\sx(t)=w_\sx(t)$.

Similarly, the estimator (\ref{eq:estimator-of-allan}) is written as 
\begin{equation}
\sigma_P^2(\tau)=\mathbb{E}\left\{\frac{1}{T}\int_{0}^{T}
\Bigl[\sy(t)\ast\mathfrak{h}_{\sy}(t)\Bigr]^{2}\;dt \right\},
\end{equation}
where 
\begin{equation}
\mathfrak{h}_{\sy}(t)=
\frac{3\sqrt{2}\,t}{\tau^{3}}\left(|t|-\tau\right)\chi_{(-\tau,\tau)}(t)  \label{eq:defhPy}
\end{equation}
is the convolution kernel which applies to $\sy(t)$.

Thanks to the fact that $\sy(t)=\dot{\sx}(t)$, $\sigma^2_P(\tau)$ can also be expressed as the running average
\begin{equation*}
\sigma_P^2(\tau)=\mathbb{E}\Biggl\{\Biggl(\int_{\mathbb{R}}\sy(s)\,w_{\sy}(s-t)\;ds\Biggr) ^{2}\Biggr\},
\end{equation*}
where 
\begin{equation*}
w_{\sy}(t)=-\frac{3\sqrt{2}\,t}{\tau^{3}}\Bigl(|t|-\tau\Bigr)\:\chi_{(-\tau,\tau)}(t)
\end{equation*}
is the weight function. Since $w_{\sy}(t)$ is odd function, it hold that $\mathfrak{h}_{\sy}(t)=-w_{\sy}(t)$. Moreover, the parabolic shape of the PVAR wavelet comes from the $t \cdot |t|$ factor in $w_{\sy}(t)$ and $\mathfrak{h}_{\sy}(t)$.

For the purpose of operation with the Fourier transform, it is convenient
to restate these expression in terms of filter or convolution
\begin{equation}
\sigma _{P}^{2}(\tau )=\mathbb{E}\{(\sy\ast \mathfrak{h}_{\sy%
})^{2}\}=\mathbb{E}\{(\sx\ast \mathfrak{h}_{\sx})^{2}\}
\label{eq:defPVARh}
\end{equation}

\begin{figure}[t]\centering
\includegraphics[width=70mm]{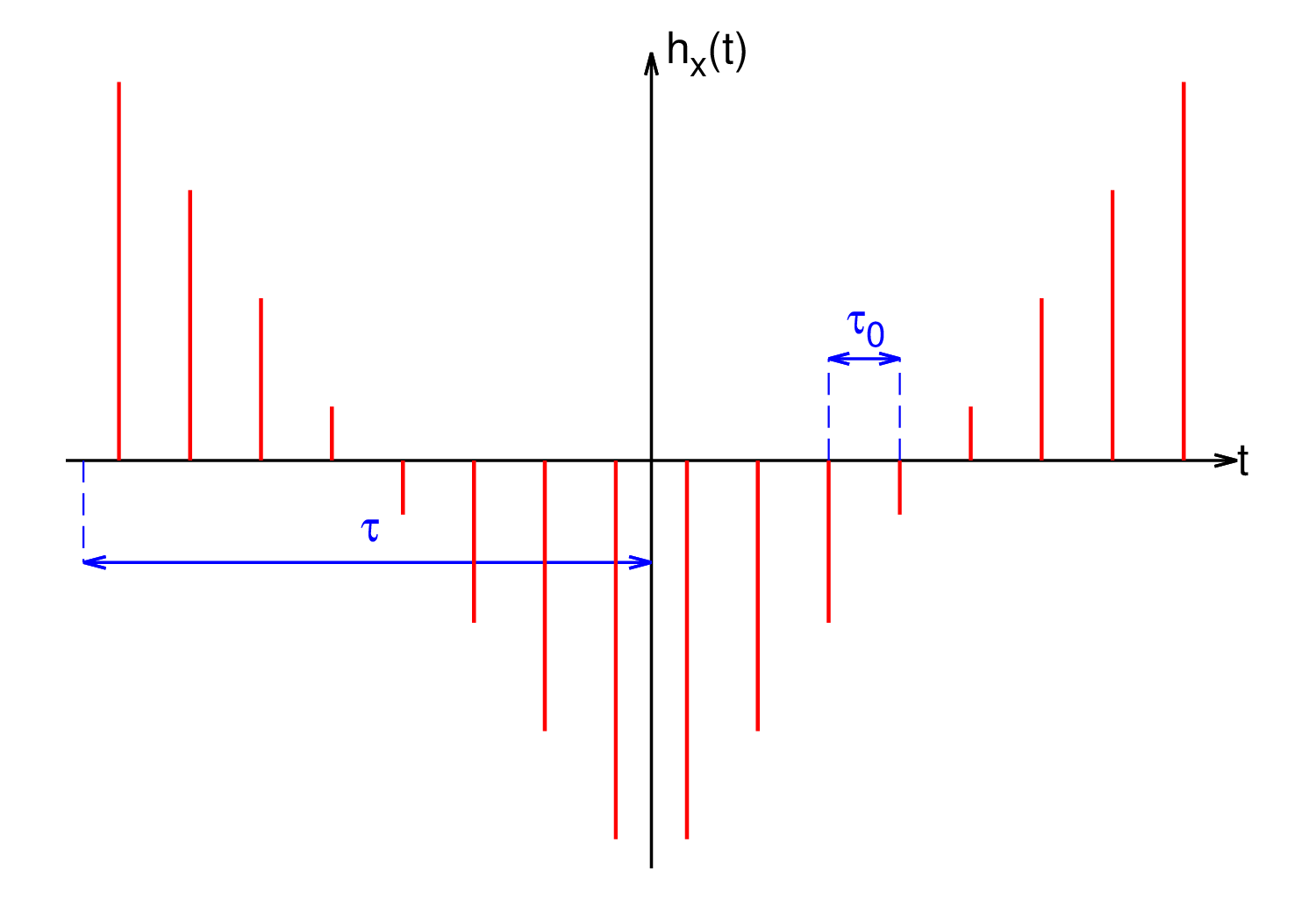}\\
\includegraphics[width=70mm]{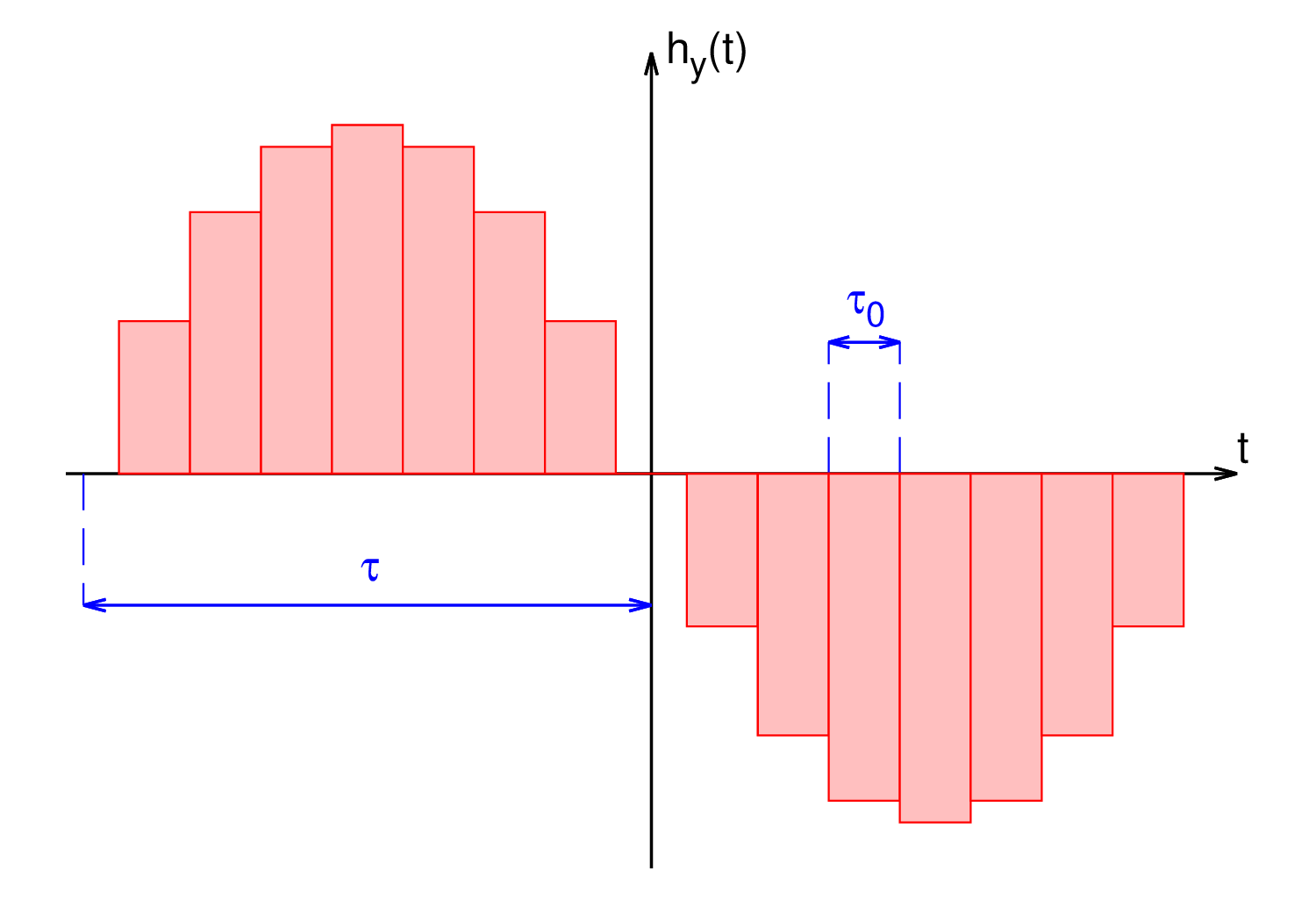}
\caption{Convolution kernels of PVAR from phase data (above) or
frequency deviations (below) for $\tau=8\tau_0$.}
\label{fig:calseq}
\end{figure}

The weight functions $w_\sx(t)$ and $w_\sy(t)$, and also the kernels $\mathfrak{h}_{\sx}(t)$ and $\mathfrak{h}_{\sy}(t)$, match the definition of power-type wavelet given in the introduction.  As a consequence of the property $\sy(t)=\dot{\sx}(t)$, it holds that $\mathfrak{h}_{\sx}(t)=\dot{\mathfrak{h}}_{\sy}(t)$.
Figure \ref{fig:calseq} shows the convolution kernels associated to PVAR.

\begin{figure}[t]\centering
\includegraphics[width=60mm]{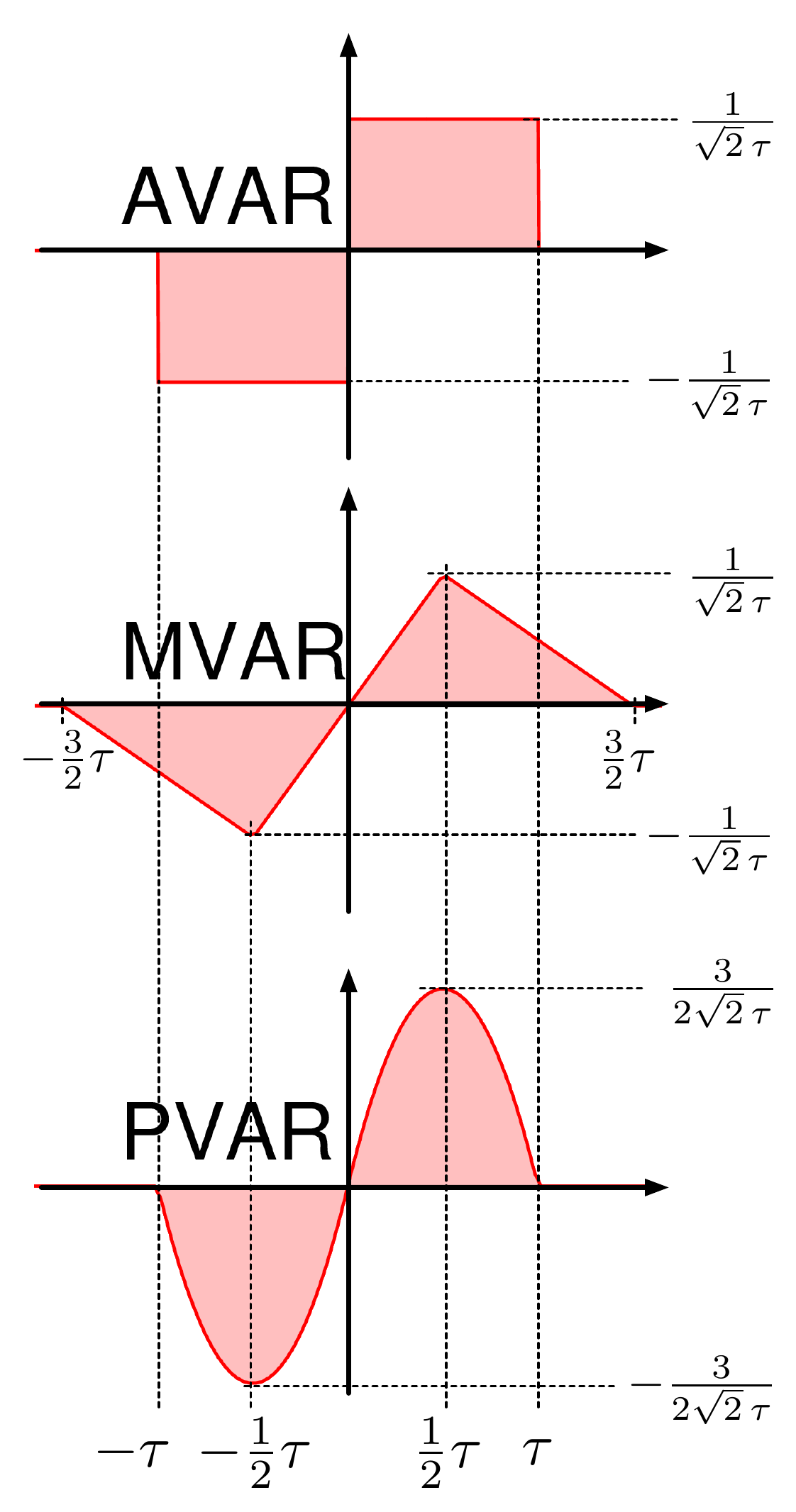}
\caption{The wavelets associated to AVAR, MVAR, and PVAR.}
\label{fig:wavelets}
\end{figure}

It is worth pointing out that our formulation is is general, as it applies to AVAR, MVAR, PVAR, and to other similar variances as well.  Of course, the wavelet depends on the counter (Fig~\ref{fig:wavelets}).

\subsection{Practical Evaluation of PVAR}
Denoting the discrete time with $\sx_{i}=\sx(i\tau _{0})$, the estimate of the two-sample variance is \cite{lesage1973}
\begin{equation}
\alpha_i=\frac{1}{\sqrt{2}\,\tau }\Bigl(-\sx_{i}+2\sx_{i+m}-\sx_{i+2m}\Bigr)
\end{equation}
for AVAR, with $M=N-2m$, and 
\begin{equation}
\alpha_{i}=\frac{1}{\sqrt{2}\,m\tau}\sum_{k =0}^{m-1}\Bigl(-\sx_{i+k}+2\sx_{i+m+k }-\sx_{i+2m+k }\Bigr)
\end{equation}
for MVAR, with $M=N-3m+1$.

Now we calculate $\alpha_{i}$ for PVAR\@.  
First, the discrete form of $\hat{\mathsf{y}}$ can be obtained from (\ref{eq:yest}) by replacing the time integral with a sum with a time increment equal to $\tau_0$.  Accordingly, $\tau$ is replaced with $m\tau_0$, $s$ with $k\tau_0$, $t$
with $i\tau_0$, and $\sx(t)$ with $\sx_{i}$
\begin{align*}
\hat{\sy}_i&=\frac{12}{m^3\tau_0^3}\sum_{k=0}^{m-1} \left(\frac{(m-1)\tau_0}{2}-k\tau_0\right)\sx_{i-k}\tau_0\\
&=\frac{12}{m^2\tau}
\sum_{k=0}^{m-1} \left(\frac{m-1}{2}-k\right)\sx_{i-k}
\end{align*}
Similarly,
\begin{equation*}
\hat{\sy}_{i+1}=\frac{12}{m^2\tau}\sum_{k=0}^{m-1} \left(\frac{m-1}{2}%
-k\right)\sx_{i+m-k}
\end{equation*}
and consequently
\begin{equation*}
\hat{\sy}_i-\hat{\sy}_{i+1}=\frac{12}{m^2\tau}\sum_{k=0}^{m-1}
\left(\frac{m-1}{2}-k\right)\bigl(\sx_{i-k}-\sx_{i+m-k}\bigr).
\end{equation*}
Second, we recall that $\sx_i$ is defined for $i\geq 0$.  Hence, we have to shift the origin by $m-1$, so that also $\hat{\sy}_i$ is defined with $i=0$
\begin{equation*}
\hat{\sy}_i-\hat{\sy}_{i+1}=\frac{12}{m^2\tau}\sum_{k=0}^{m-1}
\left(\frac{m-1}{2}-k\right)\bigl(\sx_{i+m-1-k}-\sx_{i+2m-1-k}\bigr).
\end{equation*}
Third, since the coefficient $(m-1)/2-k$ is symmetrical for $k$ and for $m-1-k$, we interchange $i+m-1-k$ with $i+m-1-(m-1-k)=i+k$, and $i+2m-1-k$ with $i+2m-1-(m-1-k)=i+m+k$
\begin{equation*}
\hat{\sy}_i-\hat{\sy}_{i+1}=\frac{12}{m^2\tau}\sum_{k=0}^{m-1}
\left(\frac{m-1}{2}-k\right)\bigl(\sx_{i+k}-\sx_{i+m+k}\bigr).
\end{equation*}
Finally, it comes
\begin{eqnarray}
\alpha_{i}&=&\frac{6\sqrt{2}}{m^{2}\tau }\sum_{k =0}^{m-1}\left(\frac{m-1}{2}-k\right) \bigl(\sx_{i+k }-\sx_{i+m+k }\bigr)
\qquad\text{for $m\ge2$}
\label{eq:kernel_omega}\\[1ex]
M&=&N-2m+2.\nonumber
\end{eqnarray}
For consistency with AVAR and MVAR, we require $\sigma^2_P(\tau_0)=\sigma^2_A(\tau_0)=\sigma^2_M(\tau_0)$, i.e. all variances are equal at sampling time $\tau_0$.  Since (\ref{eq:kernel_omega}) gives $\alpha_i=0$ for $m=1$, we redefine 
\begin{eqnarray}
\alpha_i&=&\frac{1}{\sqrt{2}\tau_0}\Bigl(-\sx_{i}+2\sx_{i+1}-
\sx_{i+2}\Bigr)
\qquad\text{for $m=1$}\\[1ex]
M&=&N-2\nonumber
\end{eqnarray}

Having $N$ samples $\{\sx_{i}\}$ taken at the interval $\tau $, 
the estimate $\hat{\sigma}_{P}^{2}(\tau )$ can be computed using
(\ref{eq:var_gen}) and (\ref{eq:kernel_omega}) as
\begin{equation}
\hat{\sigma}_{P}^{2}(\tau )=\frac{72}{Mm^{4}\tau ^{2}}\sum_{i=0}^{M-1}\left[
\sum_{k=0}^{m-1}\left(\frac{m-1}{2}-k\right)
\bigl(\sx_{i+k}-\sx_{i+m+k}\bigr) \right]^{2}.
\label{eq:defPVARx}
\end{equation}%

\subsection{Time-Domain Response}
From (\ref{eq:defPVARx}) it comes%
\begin{eqnarray}
\sigma _{P}^{2}(\tau)
&=&\mathbb{E}\left\{\hat{\sigma}_{P}^{2}(\tau )\right\}\nonumber\\
&=&\frac{72}{m^{4}\tau ^{2}}\mathbb{E}\left\{ \frac{1}{M}\sum_{i=0}^{M-1}\left[
\sum_{k=0}^{m-1}\left( \frac{m-1}{2}-k\right) \left( \sx_{i+k}-%
\sx_{i+m+k}\right) \right] \right.
\nonumber\\
&&\left.\times\left[\sum_{l=0}^{m-1}\left(\frac{m-1}{2}-l\right)
\bigl(\sx_{i+l}-\sx_{i+m+l}\bigr) \right] \right\}\nonumber\\[1em]
&=&\frac{72}{m^{4}\tau ^{2}}\sum_{k=0}^{m-1}\sum_{l=0}^{m-1}\left( \frac{m-1}{2}-k\right) \left( \frac{m-1}{2}-l\right)\nonumber\\
&&\times\Bigl[R_{\sx}((k-l)\tau_{0})-R_{\sx}((k-m-l)\tau_{0})-R_{\sx}((m+k-l)\tau_{0})+R_{\sx}((k-l)\tau_{0})\Bigr]\label{PVAR time domain}
\end{eqnarray}
where $R_{\sx}(\theta)=\mathbb{E}\left\{\sx(t)\sx%
(t+\theta)\right\}$ is the autocorrelation function of $\sx(t)$.  The autocorrelation function is detailed in Section \ref{sec:numcomp}. 
Whereas $R_\sx(\tau)$ depends on $f_L$ and $f_H$, these parameters cancel in the derivation of PVAR.

\section{Frequency Domain Representation}

\subsection{Transfer Function}

\begin{figure}[tbp]
\centering
\includegraphics[width=80mm]{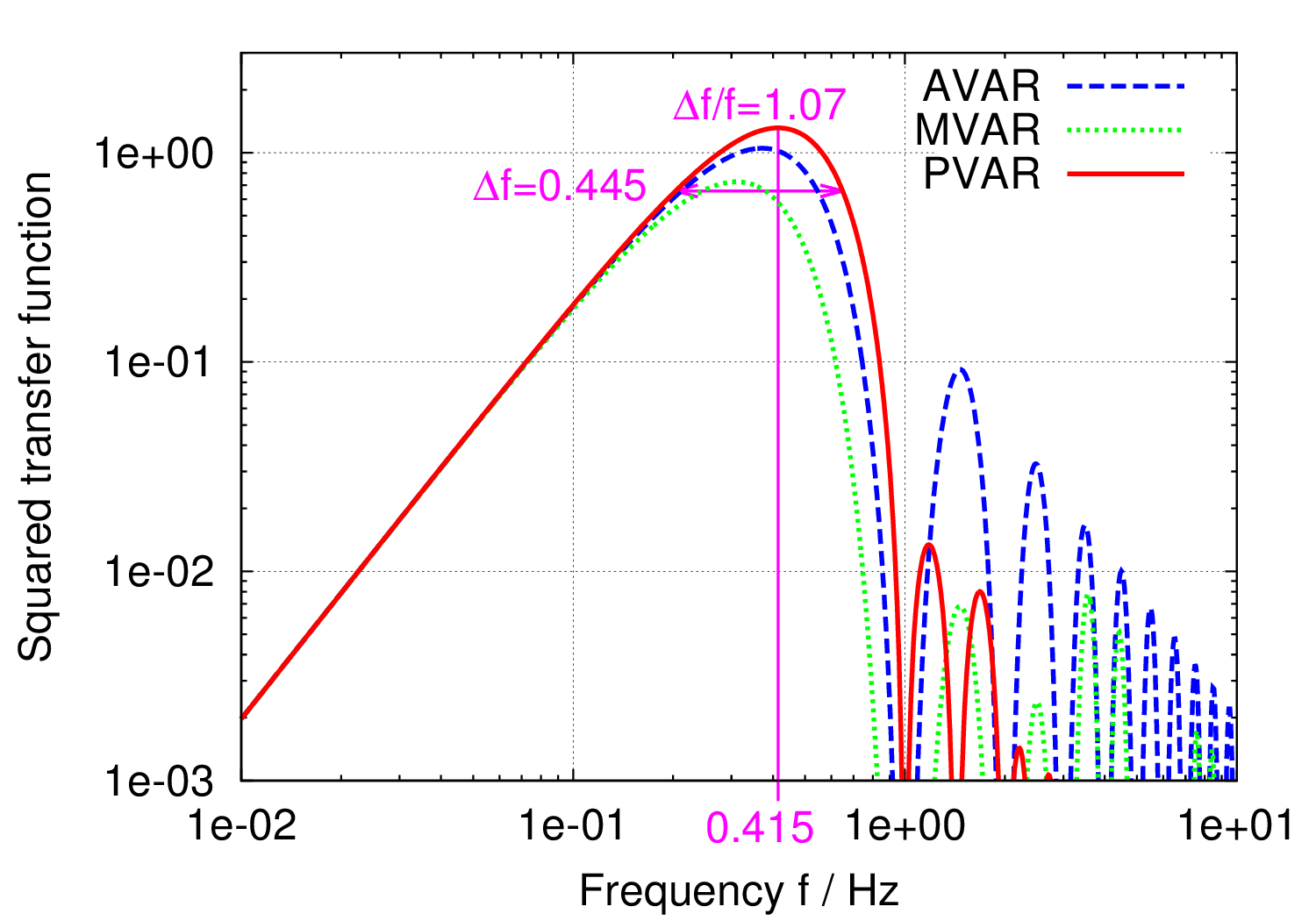}
\caption{PVAR transfer function, compared to AVAR and MVAR, for integration time is $\tau=1$ s and sampling interval $\tau_0=\tau/4=250$ ms.}
\label{fig:transfun}
\end{figure}

The transfer function $H_P(f)$ of PVAR is the Fourier transform
of the kernel $\mathfrak{h}_{\sy}(t)$. The square of its modulus is
given by
\begin{equation}
\left|H_P(f)\right|^2=\frac{9\left[2\sin^2(\pi
\tau f)-\pi\tau f\sin(2\pi\tau f)\right]^2}{2(\pi\tau f)^6}.
\label{eq:trafun}
\end{equation}%
Figure~\ref{fig:transfun} shows $\left|H_P(f)\right|^{2}$, together with the transfer function of AVAR and MVAR\@.  All are bandpass functions with approximately one octave bandwidth.  However, PVAR exhibits significantly smaller side lobes because the weight function is smoother.  This is well known with the taper (window) functions used in the digital Fourier transform \cite{Brigham:fft}.

This can be proved as follows.  The transfer function is obtained after Fourier transformation, using the property that $\mathfrak{h}_{\sy}$ is odd function
\begin{align*}
H_{\sy}(f)
&=\int_{\mathbb{R}}\mathfrak{h}_\sy(t)e^{-2i\pi ft}dt\\[1ex]
&=\frac{12i}{\sqrt{2}\tau ^{3}}\Im\left\{\int_0^\tau t\left(t-\tau
\right) e^{-2i\pi ft}\; dt\right\}.
\end{align*}
The primitive is calculated by parts integration 
\begin{align*}
\int_{0}^{\tau }t\left(t-\tau\right) e^{-2i\pi ft}dt
&=\frac{1}{4\pi^{3}f^{3}}\Bigl( \pi \tau f-ie^{-2i\pi \tau f}+\pi \tau fe^{-2i\pi \tau
f}+i\Bigr)\,.
\end{align*}%
Then,%
\begin{equation*}
H_P(f)=\frac{3i}{\sqrt{2}\pi ^{3}\tau ^{3}f^{3}}
\Bigl[1-\cos\left(2\pi\tau f\right)-\pi\tau f\sin\left(2\pi\tau
f\right) \Bigr].
\end{equation*}%
Finally, using $1-\cos\left(2\pi\tau f\right)=2\sin^2\left(\pi\tau f\right) $, we get
\begin{equation*}
H_P(f)=\frac{3i}{\sqrt{2}\pi^{3}\tau^{3}f^{3}}\Bigl[2\sin
^{2}\left( \pi \tau f\right) -\pi \tau f\sin\left(2\pi \tau f\right)
\Bigr],
\end{equation*}%
and
\begin{equation*}
|H_P(f)|^{2}=\frac{9}{2\pi ^{6}\tau ^{6}f^{6}}
\Bigl[1-\cos\left(2\pi\tau f\right)-\pi\tau f\sin\left(2\pi\tau f\right) \Bigr]^2.
\end{equation*}%

\subsection{Convergence Properties}
For small $f$, it holds that 
\begin{gather*}
\sin(\pi\tau f) \approx \pi\tau f-\frac{1}{6}\pi^3\tau
^3f^3+O(f^5)\\
\sin(2\pi\tau f) \approx 2\pi\tau f-\frac{4}{3}\pi^3\tau^3f^3+O(f^5)
\end{gather*}
so
\begin{eqnarray*}
2\sin ^{2}\left( \pi \tau f\right) -\pi \tau f\sin \left( 2\pi \tau f\right)
&\approx &2\left( \pi \tau f-\frac{1}{6}\pi ^{3}\tau ^{3}f^{3}\right)
^{2}-\pi \tau f\left( 2\pi \tau f-\frac{4}{3}\pi ^{3}\tau ^{3}f^{3}\right)
+f^{5}O(f) \\
&\approx &\frac{1}{18}\pi ^{4}\tau ^{4}f^{4}\left( \pi ^{2}\tau
^{2}f^{2}+12\right) +f^{5}O(f)
\end{eqnarray*}
then, at low frequency,
\begin{equation*}
H_P(f)\approx \sqrt{2}i\pi \tau f.
\end{equation*}
We conclude that
\begin{align*}
|H_P(f)|^2 &\approx 2\pi^2\tau^2f^2 \quad \text{at low frequency}, 
\end{align*}
thus PVAR converges for $1/f^2$ FM noise.  Similarly
\begin{align*}
|H_P(f)|^2 &\propto(\pi\tau f)^{-4} \quad\text{at high frequency},
\end{align*}
therefore PVAR converges for $f^2$ FM noise.

\subsection{Calculation of PVAR from Spectral Data}
\begin{table*}[tbp]\centering
\caption{Response of AVAR, MVAR and PVAR to the common noise types, and to drift.}
\label{tab:normvar}
\begin{tabular}{|c|l||c|c|c||c|}
\hline
& &  &  &  &  \\
Noise &$S_\sy(f)$ & AVAR $\sigma^2_A(\tau)$ & MVAR $\sigma^2_M(\tau)$ & PVAR $\sigma^2_P(\tau)$ & $\displaystyle\frac{\sigma^2_P(\tau)}{\sigma^2_M(\tau)}$ \\
type & &  &  &  &  \\ \hline\hline
& &  &  &  &  \\
White & $\sh_{2}f^{2}$ & $\displaystyle\frac{3\sh_{2}}{8\pi^2\tau_0\tau^2}$ & $%
\displaystyle\frac{3\sh_{2}}{8\pi^2\tau^3}$ & $\displaystyle\frac{3\sh_{2}}{%
2\pi^2\tau^3}$ & 4 \\
PM & & &&&  \\ \hline
& &  &  &  &  \\
Flicker & $\sh_{1}f^{1}$ & $\displaystyle\frac{\left[1.038+3\ln(\pi\tau/\tau_0)\right]\sh_1%
}{4\pi^2\tau^2}$ & $\displaystyle\frac{\left[24\ln(2)-9\ln(3)\right]\sh_{1}}{%
8 \pi^2 \tau^2}$ & $\displaystyle\frac{3\left[\ln(16)-1\right]\sh_{1}}{%

2\pi^2\tau^2}$ & 3.2 \\
PM& &  &  &  &  \\ \hline
& &  &  &  &  \\
White &$\sh_0f^0$ & $\displaystyle\frac{\sh_0}{2\tau}$ & $\displaystyle\frac{\sh_0}{4\tau}

$ & $\displaystyle\frac{3\sh_0}{5\tau}$ & 2.4 \\
FM & &  &  &  &  \\ \hline
& &  &  &  &  \\
Flicker &$\sh_{-1}f^{-1}$ & $2\ln(2)\sh_{-1}$ & $\displaystyle\frac{\left[%
27\ln(3)-32\ln(2)\right]\sh_{-1}}{8}$ & $\displaystyle\frac{2\left[7-\ln(16)%
\right]\sh_{-1}}{5}$ & 1.8 \\
FM & &  &  &  &  \\ \hline
& &  &  &  &  \\
Random &$\sh_{-2}f^{-2}$ & $\displaystyle\frac{2\pi^2\sh_{-2}\tau}{3}$ & $\displaystyle%
\frac{11\pi^2\sh_{-2}\tau}{20}$ & $\displaystyle\frac{26\pi^2 \sh_{-2}\tau}{35}$
& 1.4 \\
walk FM& &  &  &  &  \\ \hline \hline
& &  &  &  &  \\
Drift & $\sby(t)=\mathsf{D_y}t$ & $\displaystyle\frac{1}{2}\mathsf{D}_\sy^2\tau^2$ & $\displaystyle\frac{1%
}{2}\mathsf{D}_\sy^2\tau^2$ & $\displaystyle\frac{1}{2}\mathsf{D}_\sy^2\tau^2$ & 1 \\
& &  &  &  &  \\ \hline
\multicolumn{6}{|l|}{\textit{\footnotesize The lowpass cutoff frequency $f_H$, needed for AVAR, is set to $1/2\tau_0$ (Nyquist frequency)}} \\\hline
\end{tabular}%
\end{table*}

\begin{figure}[tbp]
\centering
\includegraphics[width=75mm]{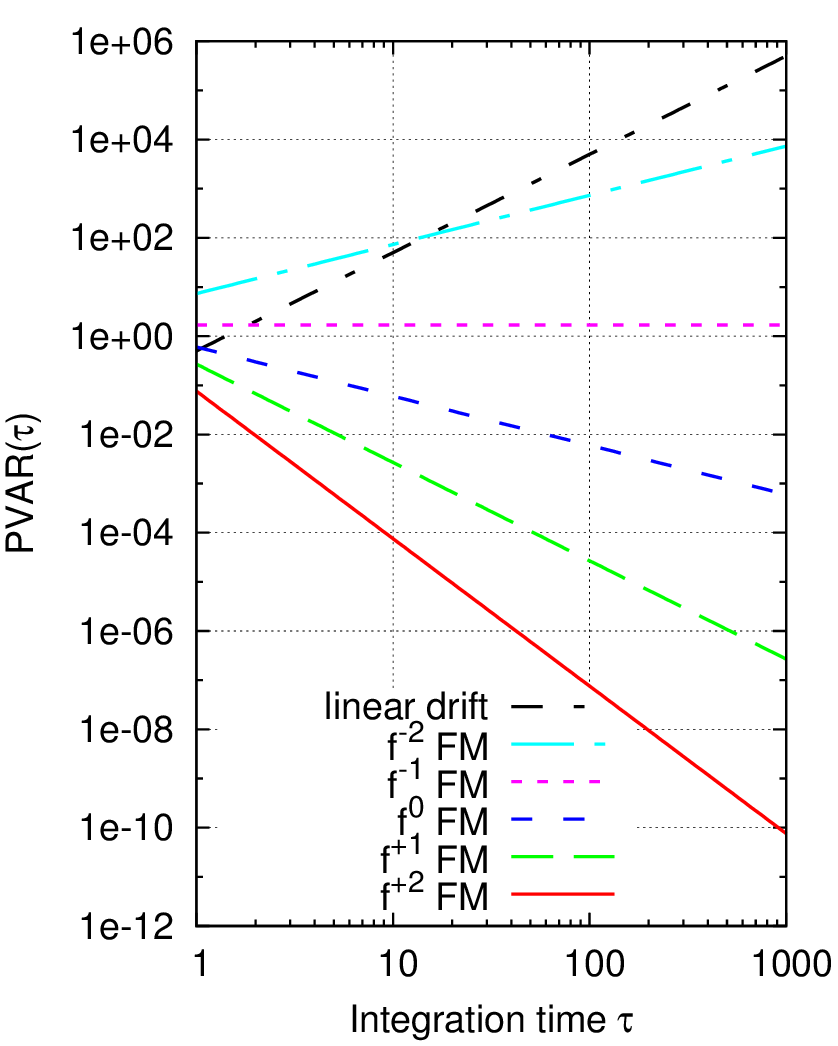}
\caption{Response of PVAR to the polynomial-law noise types, and to linear drift.}
\label{fig:respOm}
\end{figure}

Given the Power Spectral Density (PSD) $S_\sx(f)$, PVAR evaluated as
\begin{equation}
\sigma_P^2(\tau)=\int_0^\infty\left|H_P(f)\right|^2S_\sy(f)\;df.  \label{eq:OVspec}
\end{equation}
Replacing $S_\sy(f)$ with the components of the polynomial law, from $h_{-2}f^{-2}$ (random walk FM) to $h_{2}f^{2}$ (white PM), we get the
response of shown on Table \ref{tab:normvar}, together with AVAR and MVAR\@.  
Figure~\ref{fig:respOm} shows the response of PVAR to the polynomial-law noise types as a function of the integration time $\tau$.

For comparison, $\sigma^2_P(\tau)$ can also be calculated in the time domain using (\ref{PVAR time domain}), and also with Monte-Carlo simulations (see Section~\ref{sec:MCsim}). 
Time domain, frequency domain and Monte-Carlo simulations give fully consistent results.

\section{Degrees of Freedom and Confidence Interval}

\subsection{Equivalent Degrees of Freedom (EDF)}
We consider the estimates of a generic variance $\sigma^2(\tau)$, assumed $k\chi_\nu^2$-distributed, $k\in \mathbb{R}^+$.  The EDF $\nu$ depend on the integration time $\tau$, and of course on the noise type.  The mean and variance (the variance of the variance) are
\begin{equation*}
\begin{array}{rcccl}
\mathbb{E}\left\{\hat{\sigma}^2(\tau)\right\}
&=& k\mathbb{E}\left\{\chi_\nu^2\right\}
&=& k\nu \\[1ex]
\mathbb{V}\left\{\hat{\sigma}^2(\tau)\right\} 
&=& k^2\mathbb{V}\left\{\chi_\nu^2\right\}
&=& 2k^2\nu.
\end{array}
\end{equation*}
Accordingly, the degrees of freedom $\nu$ are given by
\begin{equation}
\nu=\frac{2\,\mathbb{E}\left\{\hat{\sigma}^2(\tau)\right\} ^2}{\mathbb{V}\left\{\hat{\sigma}^2(\tau)\right\} }.  \label{eq:nuest}
\end{equation}
Thus, the knowledge of $\nu$ enables to define a confidence interval around $\mathbb{E}\left\{ \hat{\sigma}^2(\tau)\right\}$ with given confidence $p$.
For applying this result to PVAR, we have then to calculate the variance of PVAR.

\subsection{Variance of PVAR}
The variance of the estimate $\hat{\sigma}^{2}(\tau )$ is given by
\begin{align}
\mathbb{V}\left\{\hat{\sigma}^2(\tau )\right\}
&=\mathbb{E}\left\{\left[\hat{\sigma}^2(\tau)-\mathbb{E}\left\{\hat{\sigma}^2(\tau)\right\}\right]^2\right\}\nonumber\\[1ex] 
&=\mathbb{E}\Biggl\{\Biggl[\frac{1}{M}\sum_{i=0}^{M-1}\alpha_i^2-\mathbb{E}\Biggl\{\frac{1}{M}\sum_{i=0}^{M-1}\alpha_i^2\Biggr\}\Biggr]^{2}\Biggr\} .
\label{eq:defvov}
\end{align}%
Expanding (\ref{eq:defvov}) yields
\begin{equation*}
\mathbb{V}\left\{\hat{\sigma}^2(\tau)\right\}=\frac{1}{M^2}\sum_{i=0}^{M-1}\sum_{j=0}^{M-1}\Bigl[\mathbb{E}\left\{ \alpha
_{i}^{2}\alpha _{j}^{2}\right\} -\mathbb{E}\left\{ \alpha _{i}^{2}\right\}
\mathbb{E}\left\{ \alpha _{j}^{2}\right\}\Bigr] .
\end{equation*}%
The Isserlis's theorem \cite{isserlis1916, isserlis1918, greenhall2015} states that, for centered and jointly Gaussian random variables $z$ and $w$
\begin{equation*}
\mathbb{E}\left\{z^2w^2\right\}-\mathbb{E}\left\{z^2\right\}-\mathbb{E}\left\{w^2\right\}=2\left[\mathbb{E}\left\{zw\right\}\right]^2.
\end{equation*}%
Assuming that \textsf{x} is a Gaussian process and that $\alpha _{i}$, $\alpha_{j}$ are two centered jointly Gaussian random variables, it comes
\begin{equation}
\mathbb{V}\left\{ \hat{\sigma}^{2}(\tau )\right\} =\frac{2}{M^{2}}%
\sum_{i=0}^{M-1}\sum_{j=0}^{M-1}\Bigl[\mathbb{E}\left\{\alpha_i\alpha_j\right\}\Bigr]^2.
\label{eq:var_var}
\end{equation}

The derivation of $\mathbb{E}\left\{\alpha_i\alpha_j\right\}$ is given in the next Section.

\subsection{Equivalent Degrees of Freedom of PVAR}
From (\ref{eq:kernel_omega}), we can calculate
\begin{align*}
\mathbb{E}\left\{\alpha_i\alpha_j\right\}
&=\frac{72}{m^4\tau^2}\mathbb{E}\left\{
\left[\sum_{k=0}^{m-1}\left(\frac{m-1}{2}-k\right)
\bigl(\sx_{i+k}-\sx_{i+m+k}\bigr)\right]\right.\\
&\quad\left.\times\left[\sum_{l=0}^{m-1}\left(\frac{m-1}{2}-l\right)
\bigl(\sx_{j+l}-\sx_{j+m+l}\bigr) \right] \right\}
\end{align*}
which expands as
\begin{align}
\mathbb{E}\left\{\alpha_i\alpha_j\right\} 
&=\frac{72}{m^4\tau^2}
\sum_{k=0}^{m-1}\sum_{l=0}^{m-1}
\left(\frac{m-1}{2}-k\right)\left(\frac{m-1}{2}-l\right)
\Bigl\{2R_{\sx}[(i+k-j-l)\tau_{0}]\Bigr. \nonumber
\\[1ex]
&\quad\Bigl.-R_{\sx}[(i+k-j-m-l)\tau_0]-R_{\sx}[(i+m+k-j-l)\tau_{0}]\Bigr\}.
\label{eq:eaiaj}
\end{align}
Thanks to (\ref{eq:var_var}) and (\ref{eq:eaiaj}), we can calculate the variance of PVAR from the autocorrelation function $R_\sx(\tau)$. For example, in the case of white PM noise we find
\begin{equation}
\mathbb{V}\left\{\hat{\sigma}_P^2(\tau)\right\} =\frac{9\sh_2^2}{70\pi^4\tau^6}
\left[23\frac{m}{M}-12\left(\frac{m}{M}\right)^2-175\frac{m}{M^2}\right]
\label{eq:vgendef}
\end{equation}%
and
\begin{equation}
\nu =\frac{35}{23m/M-12(m/M)^{2}-175m/M^{2}}.  \label{eq:nugen}
\end{equation}

\subsection{Numerical Evaluation of the EDF}\label{sec:numcomp}
The EDF can be evaluated by substituting (\ref{eq:eaiaj}) into (\ref{eq:var_var}), and then (\ref{eq:var_var}) into (\ref{eq:nuest}).  
In turn, thanks to the Wiener Khinchin theorem, stationary ergodic processes states that $R_\sx(\tau)$ can be obtained as the inverse Fourier transform of the Power Spectral Density (PSD).
Since the PSD is real and even \cite{vernotte2001,vernotte2015}, we get
\begin{equation}
R_{\sx}(\tau )=\int_{0}^{+\infty }S_{\sx}(f)\cos (2\pi\tau f)\;df.
\end{equation}%
Replacing $S_\sx(f)$ with the polynomial law from white
PM to random walk FM ($f^{-4}$ PM), we get the results shown in Table~\ref{tab:Rx}.  The derivation is rather mechanical, and done by a symbolic algebra application (Wolfram Mathematica).  For numerical evaluation --- unless the reader understanding the computer code in depth --- we recommend the approximations 
$\lim_{\rightarrow0}\mathrm{Ci}(x)=C+\ln(x)-x^2/96$, where $C\approx0.5772$ is the Euler-Mascheroni constant, 
$\lim_{x\rightarrow\infty}\mathrm{Ci}(x)=0$,
$\lim_{x\rightarrow0}\mathrm{Si}(x)=x-x^3/9$,
and $\lim_{x\rightarrow\infty}\mathrm{Si}(x)=\pi/2$.

\begin{table*}[tbp]
\caption{Autocorrelation function of the phase-time fluctuation.} 
\label{tab:Rx}
\centering
\begin{tabular}{|c|c|c|}
\hline
$S_\sx(f)$ & $R_\sx(0)$ & $R_\sx(\tau)$ \hspace{3mm}
(for $\tau\neq 0$) \\ \hline
$\sk_0$ & $\sk_0 f_H$ & 0 \\
&  &  \\
$\sk_{-1}f^{-1}$ & $\displaystyle \sk_{-1}\left[\frac{1}{2}+\ln(f_H/f_L)\right]$
& $\displaystyle \sk_{-1}\left[\frac{\cos(2\pi f_L\tau)-1+2\pi
f_L\tau\sin(2\pi f_L\tau)}{(2\pi f_L\tau)^2}\right.$ \\
&  & $\left.\vphantom{\displaystyle\frac{\cos(2\pi f_L\tau)-1+2\pi
f_L\tau\sin(2\pi f_L\tau)}{(2\pi f_L\tau)^2}}+\mathrm{Ci}(2\pi\tau f_H)-%
\mathrm{Ci}(2 \pi \tau f_L)\right]$ \\
&  &  \\
$\sk_{-2}f^{-2}$ & $\displaystyle \sk_{-2}\left[\frac{1}{f_L}-\frac{1}{f_H}%
\right]$ & $\displaystyle \sk_{-2}\left\{\frac{\cos(2\pi f_L\tau)}{f_L}-\frac{%
\cos(2\pi f_H\tau)}{f_H}\right.$ \\
&  & $\displaystyle \left.+2\pi\tau\Bigl[\mathrm{Si}(2\pi f_L\tau)-\mathrm{Si%
}(2\pi f_H\tau)\Bigr]\vphantom{\frac{\cos(2\pi f_L\tau)}{f_L}}\right\}$ \\
&  &  \\
$\sk_{-3}f^{-3}$ & $\displaystyle \frac{\sk_{-3}}{2}\left[\frac{1}{f_L^2}-\frac{1%
}{f_H^2}\right]$ & $\displaystyle \sk_{-3}\left\{\frac{\cos(2\pi f_L\tau)}{%
2f_L^2}-\frac{\cos(2\pi f_H\tau)}{2f_H^2}\right.$ \\[1ex]
&  & $\displaystyle +2\pi^2\tau^2\Bigl[\mathrm{Ci}(2\pi f_L\tau)-\mathrm{Ci}(2\pi f_H\tau)\Bigr]$ \\[2ex]
&  & $\displaystyle \left.+\pi\tau\left[\frac{\sin(2\pi f_H\tau)}{f_H}-\frac{%
\sin(2\pi f_L\tau)}{f_L}\right]\right\}$ \\
&  &  \\
$\sk_{-4}f^{-4}$ & $\displaystyle \frac{\sk_{-4}}{3}\left[\frac{1}{f_L^3}-\frac{1%
}{f_H^3}\right]$ & $\displaystyle \sk_{-4}\left\{\frac{(2\pi^2
f_H^2\tau^2-1)\cos(2\pi f_H\tau)+\pi f_H\tau\sin(2\pi f_H\tau)}{3f_H^3}
\right.$ \\[3ex]
&  & $\displaystyle -\frac{(2\pi^2 f_L^2\tau^2-1)\cos(2\pi f_L\tau)+\pi
f_L\tau\sin(2\pi f_L\tau)}{3f_L^3}$ \\
&  & $\displaystyle \left.+\frac{4\pi^3\tau^3}{3}\left[\mathrm{Si}(2\pi
f_H\tau)-\mathrm{Si}(2\pi f_L\tau)\right]\right\}$\\
\hline
\multicolumn{3}{|l|}{$f_L$ and $f_H$ are the highpass and lowpass cutoff frequencies which set the process bandwidth}\\
\multicolumn{3}{|l|}{$\mathrm{Ci}(x)$ and $\mathrm{Si}(x)$ are the Cosine and Sine Integral functions}\\
\hline
\end{tabular}%
\end{table*}

\begin{figure}[tbp]\centering
\includegraphics[height=6.5cm]{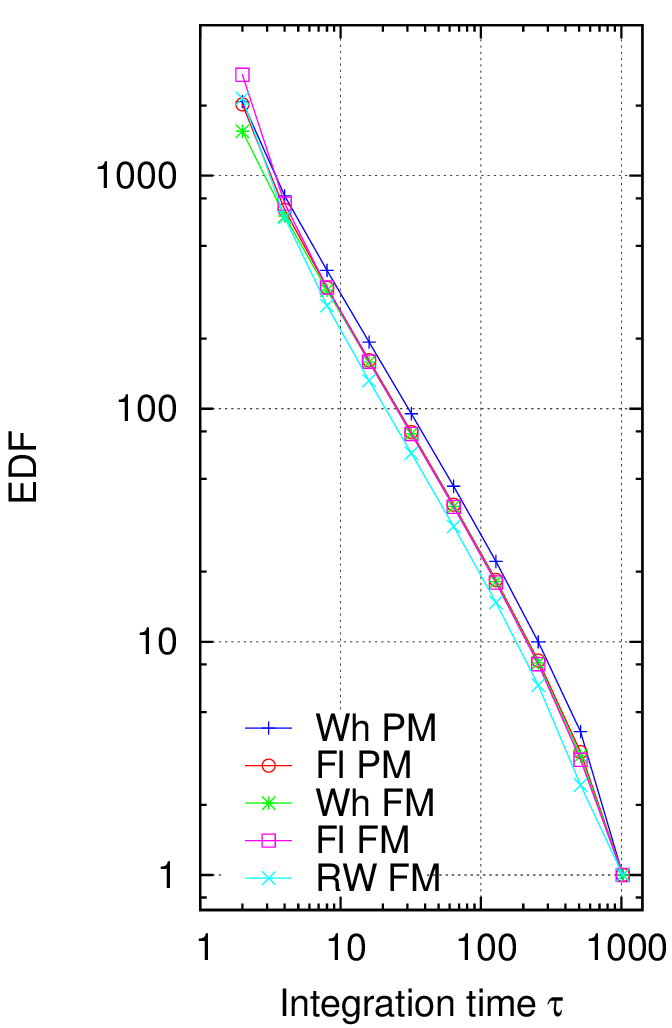}
\includegraphics[height=6.5cm]{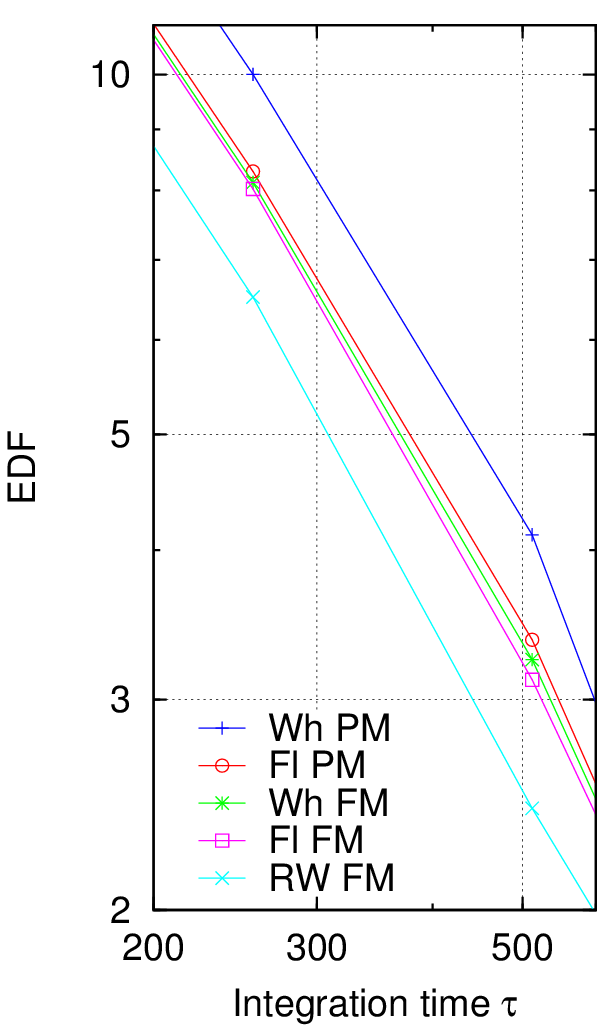}
\caption{Numerical computation of the PVAR EDF for the common noise types.  The right hand plot is a crop of the left one.}
\label{fig:edfnum}
\end{figure}

As an example, we take a data record of $N=2048$, $\tau_0=1$ s, high cut-off frequency $f_H=\frac{1}{2\tau_0}$ (equal to the Nyquist frequency), low cut-off frequency $f_L=\frac{1}{256\,N\tau_0}$ (see \cite{vernotte2015} for the physical meaning of $f_L$) and $\tau\in\left\{\tau_0,2\tau_0,4\tau_0,\ldots,1024\tau_0\right\}$.  Figure~\ref{fig:edfnum} shows the EDF for the common noise types. 
Zooming in (Fig.~\ref{fig:edfnum} right), we see that the plots do not overlap.

\subsection{Monte-Carlo simulations\label{sec:MCsim}}

Another way to assess the EDF is by simulated time series. We generated 10\,000 sequences of $N=2048$ samples for each type of noise using the ``\emph{bruiteur}'' noise simulator \cite{sigmatheta}, which is based on filtered white noise.  This code is a part of the SigmaTheta software package, available on the URL given by \cite{sigmatheta}. It has been validated by more than 20 years of intensive use at the Observatory of Besancon.
Again, the EDF are calculated using (\ref{eq:nuest}). 

\begin{figure}[tbp]
\centering
\includegraphics[width=80mm]{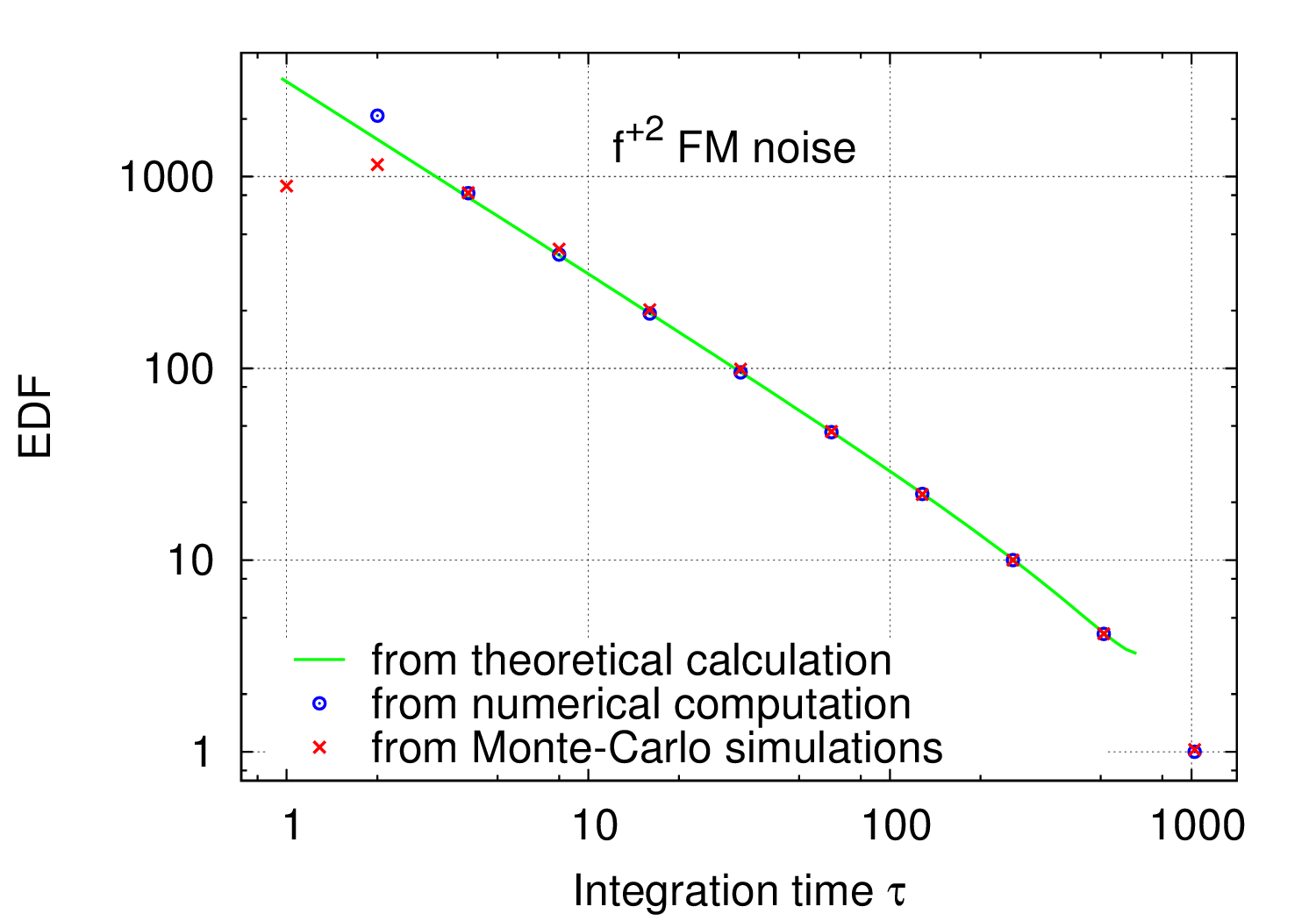}
\caption{Comparison of the EDF calculated analytically (\ref{eq:nugen}), by numerical computation (Sec.~\ref{sec:numcomp}), and assessed by Monte-Carlo simulation (Sec.~\ref{sec:MCsim}).}
\label{fig:cmpEDF}
\end{figure}

In the end, we compared three methods, the autocorrelation function $R_\sx(\tau)$ with (\ref{eq:var_var}) and (\ref{eq:eaiaj}), the Monte-Carlo simulation with \emph{bruiteur} code, and the analytical solution (\ref{eq:nugen}), the latter only with white noise.  Figure~\ref{fig:cmpEDF} compares the EDF obtained with these three methods.  The results match well, with a discrepancy of a few percent affecting only the first two points ($\tau\le2\tau_0$).
The reason is that, with such a small $\tau/\tau_0$ ratio, the $w_\sy$ weight function is a poor approximation of the parabola of the $\Omega$ counter (see \cite{rubiola2015omega}). 

\begin{figure*}[tbp]\centering
\includegraphics[scale=0.6]{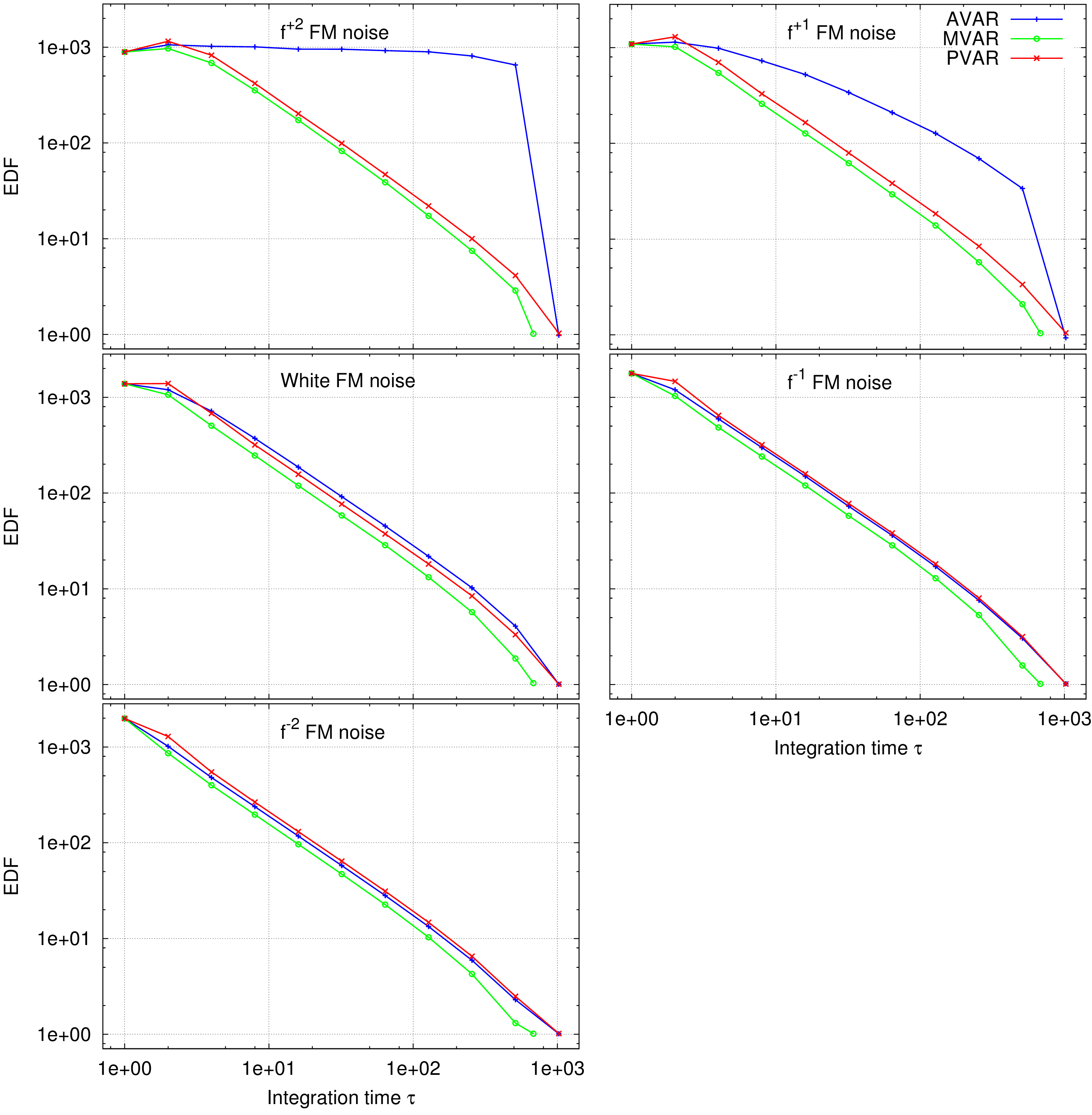}
\caption{Comparison of the Equivalent Degrees of Freedom of AVAR, MVAR and
PVAR for the different types of noise. All noise sequences were simulated
with a unity coefficient noise, 2048 samples and a sampling frequency of 1
Hz.}
\label{fig:EDFvsn}
\end{figure*}

Table~\ref{tab:EDFvsn} and Fig.~\ref{fig:EDFvsn} compare the EDF of PVAR to  AVAR and MVAR\@.  MVAR is limited to $\tau=682$ because the wavelet support (span) is $3\tau$ instead of $2\tau$.

\begin{table*}[tbp]
\caption{Comparison of the EDF of AVAR, MVAR and PVAR for the common noise types.}
\label{tab:EDFvsn}
\centering
\begin{tabular}{l|cccccccccccc}
$\tau/\tau_0$ & 1 & 2 & 4 & 8 & 16 & 32 & 64 & 128 & 256 & 512 & 682 & 1024
\\ \hline
\multicolumn{13}{l}{White PM ($f^{+2}$ FM)} \\
AVAR & 892 & 1060 & 1020 & 1010 & 955 & 953 & 922 & 896 & 811 & 652 &  &
0.981 \\
MVAR & 891 & 970 & 685 & 355 & 173 & 82.5 & 38.9 & 17.3 & 7.48 & 2.88 & 1.02
&  \\
PVAR & 892 & 1150 & 824 & 419 & 202 & 99.1 & 46.9 & 22.0 & 10.0 & 4.13 &  &
1.03 \\ \hline
\multicolumn{13}{l}{Flicker PM ($f^{+1}$ FM)} \\
AVAR & 1090 & 1140 & 984 & 728 & 523 & 340 & 209 & 127 & 69.5 & 33.8 &  &
0.930 \\
MVAR & 1090 & 1020 & 544 & 258 & 126 & 62.1 & 29.3 & 13.9 & 5.73 & 2.09 &
1.04 &  \\
PVAR & 1090 & 1300 & 701 & 329 & 165 & 79.4 & 38.2 & 18.4 & 8.42 & 3.36 &  &
1.05 \\ \hline
\multicolumn{13}{l}{White FM ($f^0$ FM)} \\
AVAR & 1380 & 1200 & 716 & 372 & 186 & 91.7 & 45.3 & 21.8 & 10.2 & 4.07 &  &
1.01 \\
MVAR & 1380 & 1060 & 505 & 247 & 119 & 58.4 & 28.6 & 13.2 & 5.71 & 1.87 &
1.04 &  \\
PVAR & 1380 & 1390 & 680 & 319 & 157 & 76.7 & 37.5 & 18.2 & 8.43 & 3.32 &  &
1.01 \\ \hline
\multicolumn{13}{l}{Flicker FM ($f^{-1}$ FM)} \\
AVAR & 1780 & 1200 & 595 & 299 & 150 & 72.8 & 36.1 & 17.1 & 7.58 & 3.05 &  &
1.02 \\
MVAR & 1780 & 1030 & 484 & 241 & 120 & 57.9 & 28.5 & 12.9 & 5.32 & 1.58 &
1.02 &  \\
PVAR & 1780 & 1470 & 648 & 319 & 159 & 77.8 & 38.2 & 18.2 & 8.01 & 3.16 &  &
1.02 \\ \hline
\multicolumn{13}{l}{Random walk FM ($f^{-2}$ FM)} \\
AVAR & 1990 & 1020 & 480 & 238 & 117 & 57.9 & 28.1 & 13.3 & 5.93 & 2.29 &  &
1.01 \\
MVAR & 1990 & 861 & 398 & 197 & 96.5 & 47.1 & 22.6 & 10.3 & 4.26 & 1.31 &
1.02 &  \\
PVAR & 1990 & 1290 & 548 & 266 & 131 & 64.3 & 31.2 & 14.8 & 6.53 & 2.49 &  &
1.02%
\end{tabular}%
\end{table*}

\section{Detection of Noise Processes}\label{sec:detection}
\begin{figure}[t]\centering
\includegraphics[width=8cm]{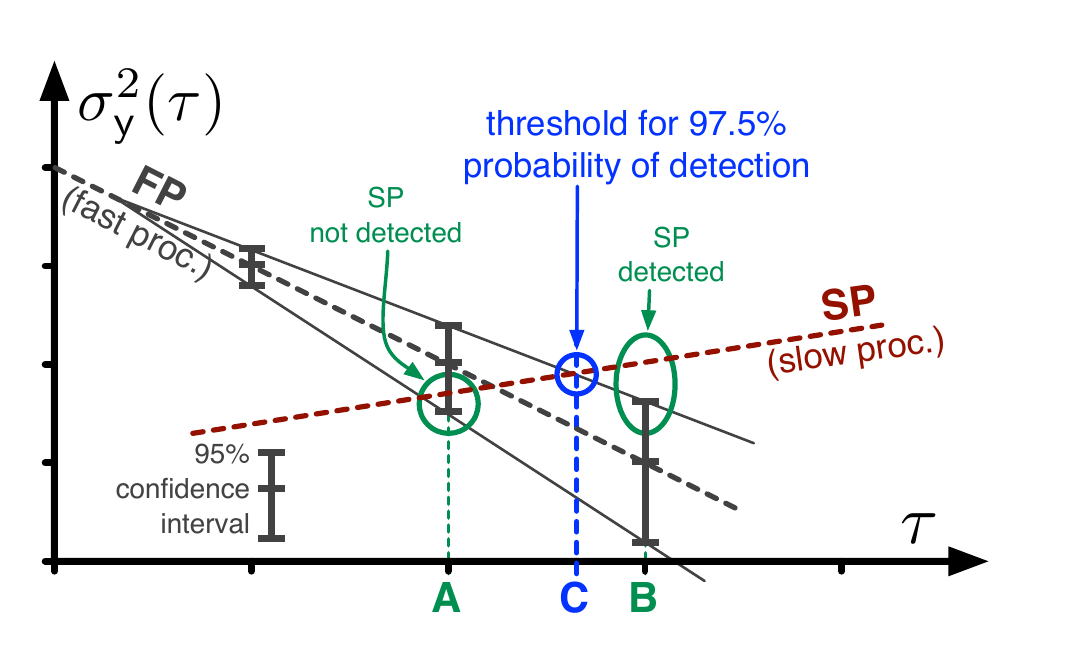}
\caption{Concept of noise process detection.  The process \textsf{SP} is barely visible in A (${\approx}50\%$ probability), detected with a probability of 97.5\% in C (threshold of nearly certain detection), and detected with ${>}97.5\%$ probability in B.}
\label{fig:Process-detection}
\end{figure}
Running an experiment, we accumulate a progressively larger number $N$ of samples $\sx_k$.  As $N$ gets larger, we fill up the $\hat{\sigma}^2_\sy(\tau)$ plot adding new points at larger $\tau$.  Besides, at smaller $\tau$ the error bars shrink because the number of degrees of freedom increases.
Looking at the log-log plot, we find the fast processes on the left and the slow processes on the right.  This is 
 due to the nearly-polynomial law $\tau^k$ of Table~\ref{tab:normvar}.  Having said that, we address the question of which variance is the most efficient tool at detecting a slower process `\textsf{SP}' in the presence of a faster process `\textsf{FP}' as illustrated in Fig.~\ref{fig:Process-detection}.
The criterion we choose is the lowest level of the \textsf{SP} that can be detected 
\begin{itemize}
\item with a probability of 97.5\% (i.e., two sigma upper bound)
\item in the presence of the faster process \textsf{FP} of given level, 
\item using a data record of given length $N$.
\end{itemize}
Our question about the most efficient tool relates to relevant practical cases detailed underneath.  

Our comparison is based on a simulation with $N=2048$ samples uniformly spaced by $\tau_0=1$ s.  So, the lowest $\tau$ is equal to 1 s, and the largest $\tau$ is equal to $N\tau_0/2=1024$ s for AVAR and PVAR, and to $N\tau_0/3=682$ s for MVAR\@.  

For fair comparison, we re-normalize the variances for the same response to the \textsf{SP} process.  For example, the response to white FM noise $S_\sy(f)=\mathsf{h}_0$ is $\mathsf{h}_0/2\tau$ for the AVAR, $\mathsf{h}_0/4\tau$ for the MVAR, and $3\mathsf{h}_0/5\tau$ for the PVAR\@.  Accordingly, a coefficient of 2, 4, or 5/3 is applied, respectively.  
Of course, this re-normalization makes sense only for comparison, and should not be used otherwise.

The results are shown in Fig.~\ref{fig:ddlf}, and discussed in Sections \ref{ssec:detection-first} to \ref{ssec:detection-last}.
Each simulation is averaged on $10^4$ runs.
All plots show AVAR (blue), MVAR (green) and PVAR (red) for the \textsf{FP} process, with the two-sigma uncertainty bars, and the \textsf{SP} process (grey).  We set the reference value of the \textsf{SP} process at the lowest level that PVAR can detect with a probability of 97.5\%, i.e., at the upper point of the two-sigma uncertainty bar at $\tau=1024$ s.  This is highlighted by a grey circle at $\tau=1024$ s.  

\def\ObsoleteFigure{%
\begin{figure*}[t]
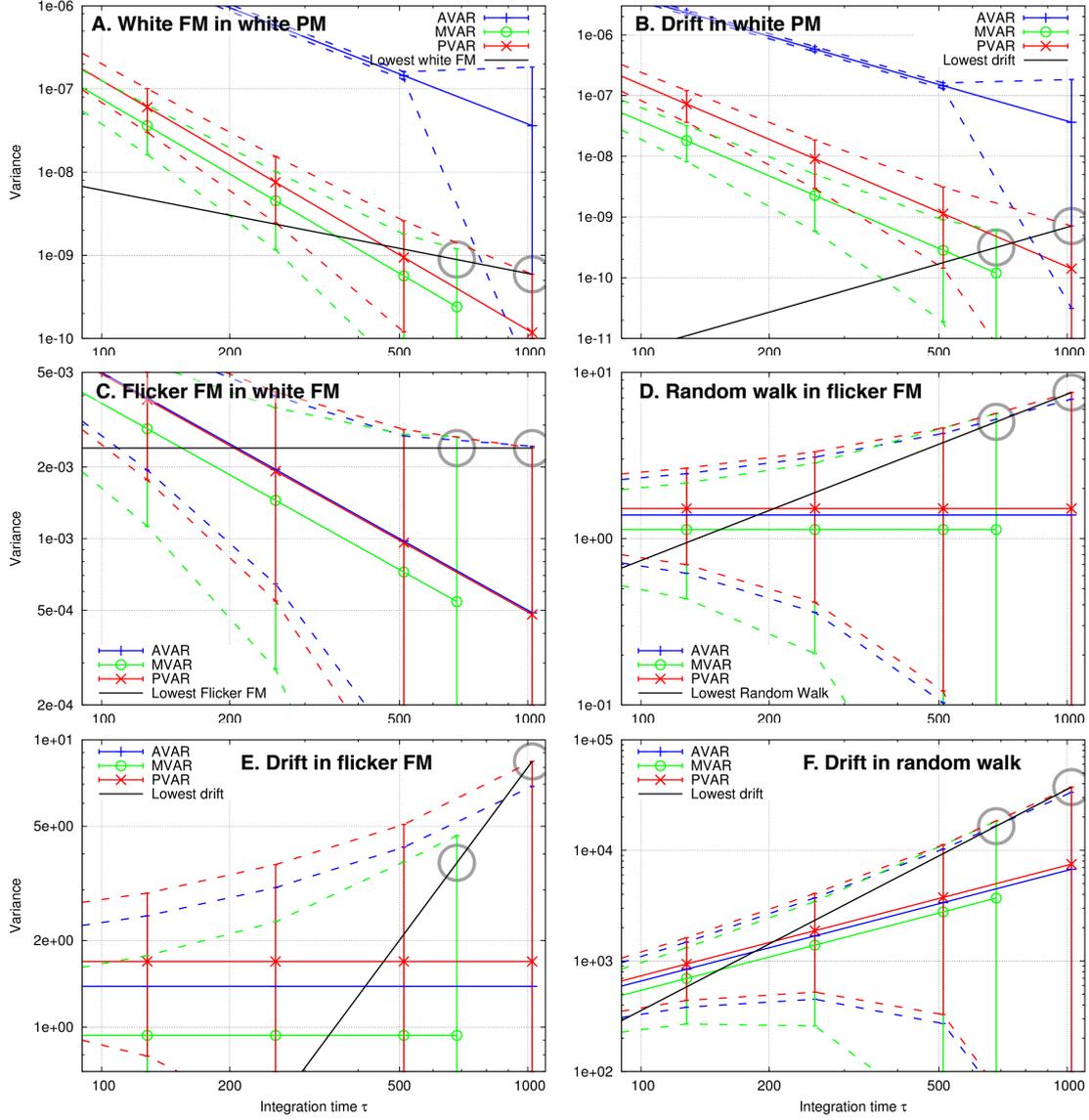
\centering
\begin{tabular}{lr}
\multicolumn{2}{c}{Short term detection:} \\ 
A. White FM in white PM & B. Drift in white PM \\ 
\includegraphics[width=7cm]{ConfInt_1e4_fp2_whFM_zoom.png} & %
\includegraphics[width=7cm]{ConfInt_1e4_fp2_zoom.png} \\ 
\multicolumn{2}{c}{Long term detection:} \\ 
C. Flicker FM in white FM & D. Random walk in flicker FM \\ 
\includegraphics[width=7cm]{detection_FlFM_vs_WhFM_zoom.png} & %
\includegraphics[width=7cm]{detection_RWFM_vs_FlFM_zoom.png} \\ 
E. Drift in flicker FM & F. Drift in random walk \\ 
\includegraphics[width=7cm]{ConfInt_fm1_zoom.png} & %
\includegraphics[width=7cm]{ConfInt_fm2_zoom.png}%
\end{tabular}%
\caption{Corner detection for the most common noise types (grey
circles). The 97.5 \% upper bounds of the confidence intervals over the variance estimates are figured by dashed lines (blue for AVAR, green for MVAR and red for PVAR). The lowest detected noise or drift by PVAR is represented as a solid black line.}
\label{fig:ddlf}
\end{figure*}}

\begin{figure*}[t]\centering
\includegraphics[scale=0.6]{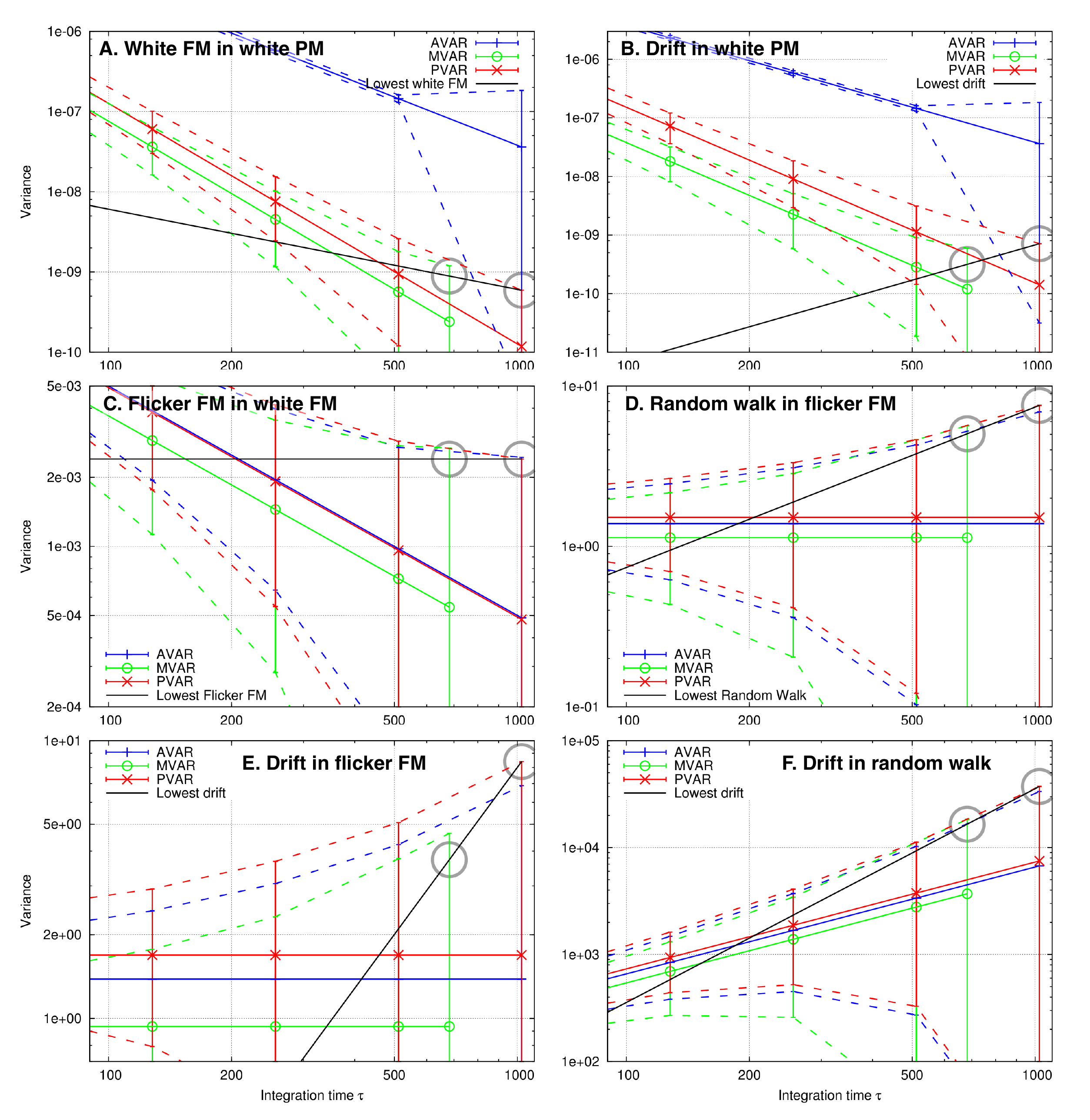}
\caption{Corner detection for the most common noise types (grey
circles). The 97.5 \% upper bounds of the confidence intervals over the variance estimates are figured by dashed lines (blue for AVAR, green for MVAR and red for PVAR). The lowest detected noise or drift by PVAR is represented as a solid black line.}
\label{fig:ddlf}
\end{figure*}

\subsection{Noise detection in the presence of white PM noise (Fig.\,\ref{fig:ddlf}\,A-B)\label{ssec:detection-first}}
White PM noise is a limiting factor in the detection of other noise processes because it is the dominant process in the front end of most instruments used to assess the frequency stability.  
We show the effect of white PM in two opposite cases, white FM noise and frequency drift.  The former is present in all atomic standards, while the latter is present in all oscillators and standards, except in the primary standards.  Frequency drift is a severe limitation in cavity stabilized lasers, and in other precision oscillators  based on the mechanical properties of an artifact.  

The classical AVAR is clearly a poor option because of its $\tau^{-2}$ response to white PM, versus the $\tau^{-3}$ of the other variances.
This is confirmed in our simulations.

It is seen on Fig.~\ref{fig:ddlf}\,A-B that in both cases MVAR cannot detect the slow process.  The lowest value of MVAR (green plot) at 97.5\% confidence (grey circle at $\tau=682$ s) exceeds the reference grey line.  

The conclusion is that PVAR exhibits the highest detection sensitivity in the in the presence of white PM noise.

\subsection{Detection of flicker FM noise in the presence of white FM noise (Fig.\,\ref{fig:ddlf}\,C)}
The detection of frequency flicker in the presence of white FM noise is a typical problem of passive atomic standards.
Such standards show white FM noise originated from the signal to noise ratio, and in turn from beam intensity, optical contrast, or other parameters depending on the physics of the standard.  Generally, after the white FM noise rolls off, $\sigma^2_\sy(\tau)$ hits the flicker floor.
Cs clocks are a special case because they do not suffer from random walk and drift.  So, flicker of frequency is the ultimate limitation to long-term stability, and in turn to timekeeping accuracy. 
In commercial standards, flicker FM exceeds the white FM at approximately 1 day integration time.  Thus, fast detection of flicker enables early estimation of the long term behavior, and provides a useful diagnostic.

We see on Figure~\ref{fig:ddlf}\,C that the three variances show similar performances, with a small superiority of AVAR and PVAR\@.  Again, MVAR suffers from the wider support of the wavelet, $3\tau$ instead of $2\tau$. AVAR has a distinguished history of beeing the favorite tool of time keepers. 

\subsection{Detection of slow phenomena (Fig.~\ref{fig:ddlf}\,D-E-F)\label{ssec:detection-last}}
It is often useful to detect the corner where random walk or drift exceed the flicker floor, or where the drift exceeds the random walk.  This problem is typical of Rb clocks and H masers, and also of precision oscillators based on mechanical properties of a resonator.
Our simulation shows that AVAR is superior, but PVAR has a detection capability close to AVAR\@.  Conversely, MVAR is 
 the poorest choice.

\section{Discussion and Conclusion}
PVAR is wavelet-like variance broadly similar to AVAR and MVAR, and intended for similar purposes.  It derives from AVAR and MVAR after replacing the $\Pi$ and $\Lambda$ counter with the $\Omega$ counter, in turn based on the linear regression on phase data \cite{rubiola2015omega}.

On closer examination, we notice that AVAR and MVAR address different problems.  In the presence of white PM noise, MVAR has a dependence as $1/\tau^3$ instead of $1/\tau^2$.  This is a good choice in microwave photonics and in other applications where the measurement of short term stability is important.
The problem with MVAR is that the wavelet spans over $3\tau$ instead of $2\tau$.  Hence, AVAR is preferred for the measurement of long term stability and in timekeeping, where the largest value of $\tau$ on the plot is severely limited by the length of the available data record. 
PVAR on the other hand is a candidate replacement for both because it features the $1/\tau^3$ dependence of MVAR and the $2\tau$ measurement time of AVAR\@.

PVAR compares favorably to MVAR because it provides larger EDF, and in turn a smaller confidence interval.
The objection that PVAR gives a larger response to the same noise level (right hand column of Table~\ref{tab:normvar}) is irrelevant because the response is just a matter of normalization.
It is only in the region of fast processes that AVAR has higher EDF than PVAR (Fig.~\ref{fig:EDFvsn}), but this happens where AVAR is certainly not the preferred option.

The best of PVAR is its power to detect and identify weak noise processes with the shortest data record.  We have seen in Sec.~\ref{sec:detection} that PVAR is superior to MVAR in all cases, and also superior to AVAR for all short-term and medium-term processes, up to flicker FM included.  
AVAR is just a little better with random walk and drift.   

In conclusion, theory and simulation suggest that PVAR is an improved replacement for MVAR in all cases, provided the computing overhead can be accepted.  
Whether or not AVAR is preferable to PVAR for timekeeping is a matter of discussion.  AVAR renders the largest $\tau$ with a given data record.  This is the case of random walk and drift.  By contrast, PVAR is superior at detecting the frequency flicker floor, which is the critical parameter of the primary frequency standards used in timekeeping.  These standards are supposed to be free from random walk and drift.
Otherwise, when rendering the largest $\tau$ is less critical, PVAR is until now the best option.

\section{Acknowledgements}
This work is supported by the ANR Programme d'Investissement d'Avenir in progress at the TF Departments of FEMTO-ST Institute and UTINAM (Oscillator IMP, First-TF and Refimeve+), and by the R\'{e}gion de Franche-Comt\'{e}.

We wish to thank Charles Greenhall for his valuable help concerning Isserlis' theorem.

\bibliography{omega.bib}
\end{document}